\documentclass[a4paper,11pt]{article}
\usepackage[utf8]{inputenc}
\pdfoutput=1 

\usepackage{jcappub} 
\usepackage{macros}

\usepackage{parskip}

\usepackage[dvipsnames]{xcolor}

\usepackage[T1]{fontenc} 

\newcommand{\rh}[1]{{#1}_{\mathrm{RH}}}
\newcommand{\gs}[1]{g_{*}(#1)}
\newcommand{\gss}[1]{g_{*S}(#1)}
\newcommand{\dis}{e^{(\epsilon + \Delta)} \pm 1}
\newcommand{\gkd}{\left[\frac{g_k}{g_{\rm dom}}\right] ^{\frac{1}{6}}}
\newcommand{\gpkd}{\left[\frac{g_{\rm pk}}{g_{\rm dom}}\right] ^{\frac{1}{6}}}
\newcommand{\beqa}{\begin{eqnarray}}
\newcommand{\eeqa}{\end{eqnarray}}

\usepackage[numbers]{natbib}
\usepackage{amsmath}
\usepackage{multirow}

\title{\boldmath The Effects Of Relativistic Hidden Sector Particles on the Matter Power Spectrum}
\author[a,b,c,1]{Himanish Ganjoo,\note{Corresponding author.}}
\author[d]{Adrienne L. Erickcek,}
\author[e]{Weikang Lin,}
\author[a,c]{and Katherine J. Mack}

\affiliation[a]{Department of Physics, North Carolina State University, Raleigh, NC 27695, USA}
\affiliation[b]{Department of Physics and Astronomy, University of Waterloo, Waterloo, ON, N2L 3G1, Canada}
\affiliation[c]{Perimeter Institute of Theoretical Physics, 31 Caroline St. N., Waterloo, ON, N2L 2Y5, Canada}
\affiliation[d]{Department of Physics and Astronomy, University of North Carolina at Chapel Hill, Phillips Hall CB 3255, Chapel Hill, NC 27599, USA}
\affiliation[e]{Tsung-Dao Lee Institute (TDLI) and School of Physics and Astronomy, Shanghai Jiao Tong University, Shengrong Road 520, 201210 Shanghai, P. R. China}

\emailAdd{hganjoo@ncsu.edu}

\abstract{
If dark matter resides in a hidden sector minimally coupled to the Standard Model, another particle within the hidden sector might dominate the energy density of the early universe temporarily, causing an early matter-dominated era (EMDE). During an EMDE, matter perturbations grow more rapidly than they would in a period of radiation domination, which leads to the formation of microhalos much earlier than they would form in standard cosmological scenarios. These microhalos boost the dark matter annihilation signal, but this boost is highly sensitive to the small-scale cut-off in the matter power spectrum. If the dark matter is sufficiently cold, this cut-off is set by the relativistic pressure of the particle that dominates the hidden sector. We determine the evolution of dark matter density perturbations in this scenario, obtaining the power spectrum at the end of the EMDE. We analyze the suppression of perturbations due to the relativistic pressure of the dominant hidden sector particle and express the cut-off scale and peak scale for which the matter power spectrum is maximized in terms of the properties of this particle. We also supply transfer functions to relate the matter power spectrum with a small-scale cut-off resulting from the pressure of the dominant hidden sector particle to the matter power spectrum that results from a cold hidden sector. These transfer functions facilitate the quick computation of accurate matter power spectra in EMDE scenarios with initially hot hidden sectors and allow us to identify which models significantly enhance the microhalo abundance. 
}

\begin{document}
\raggedbottom
\maketitle
\setlength{\abovedisplayskip}{4pt}
\setlength{\belowdisplayskip}{4pt}

\section{Introduction}

Recent null results for WIMP dark matter in direct detection experiments \cite{dmx1,dmx3,dmx2} and collider searches \cite{partx2,partx3,partx4,partx5,partx6,partx7,partx8,partx9,partx10,partx11} have prompted interest in theories in which dark matter lives in a hidden sector only weakly coupled to the Standard Model \cite{hid7,hid2,hid6,hid3,hid1,hidx}. In several hidden sector theories, long-lived massive particles dominate the energy content of the universe prior to Big Bang Nucleosynthesis (BBN), leading to an early matter-dominated era (EMDE) \cite{chen_emde,zhang,hid4,hid5,codm1,codm2,ae2020,fts-rev,cannibal_big}.  An EMDE enhances small-scale density perturbations in dark matter because subhorizon dark matter perturbations grow linearly with scale factor during matter domination, as opposed to the logarithmic growth that occurs during radiation domination \cite{emde2,emde1,emde3,ae15}. This growth can lead to the formation of dense sub-Earth-mass microhalos long before structures are expected to form in scenarios without an EMDE \cite{emde2,emde1,ae15}. 

Although these microhalos do not affect the large-scale structure of the universe, they boost dark matter (DM) annihilation rates, potentially producing detectable gamma-ray signals \cite{ae15,aew16,blanco19,sten_gr}. The DM annihilation signal is highly sensitive to the small-scale cut-off in the matter power spectrum because the cut-off scale sets the formation times and central densities of the microhalos that form due to an EMDE \cite{ae15,sten_gr, blinov}. For instance, changing the cut-off scale by a factor of two causes the DM annihilation boost to increase by two orders of magnitude \cite{sten_gr}. Therefore, an accurate calculation of this small-scale cut-off is key to observationally constraining scenarios with an EMDE. In this work, we determine the small-scale cut-off scale that results from the relativistic pressure of the particle that dominates the hidden sector.  

If the particle that dominates the energy density of the universe during the EMDE is initially relativistic, the growth of density perturbations is inhibited for modes that enter the horizon while the particle has significant pressure. We obtain exact solutions of the evolution of perturbations during an EMDE caused by a massive particle (which we call $Y$) in the hidden sector. We include the process by which this particle transitions from relativistic to nonrelativistic behavior before dominating the energy content of the universe. While perturbation equations for a relativistic hidden sector particle have been solved previously for a single set of parameters \cite{zhang}, we provide analytical expressions for the power spectrum peak and cut-off scales in terms of the statistics of the $Y$ particles and the initial ratio of densities of $Y$ and Standard Model (SM) particles.  It is also possible that the $Y$ particle experiences cannibalistic number-changing interactions that alter the evolution of its pressure; the resulting cut-off to the matter power spectrum was computed in Refs.~\cite{ae2020, cannibal_big}.  Our analysis of how the $Y$ particle generates a cut-off in the matter power spectrum in the absence of such interactions completes our understanding of how the pressure of the particle that dominates the energy density during the EMDE inhibits the growth of dark matter perturbations during the EMDE.  

We provide fitting forms for transfer functions between the cases with a hot and cold hidden sector. These transfer functions facilitate the easy computation of the power spectrum cut-off caused by the pressure suppression of density perturbations. We also consider how our transfer functions change the boost factor calculations presented in Ref. \cite{blanco19} (hereafter B19), in which the cut-off in the power spectrum was taken to be a Gaussian function of wavenumber with the cut-off scale set as the horizon scale when the mass of the dominant hidden sector particle is equal to the hidden sector temperature. Finally, we use our transfer functions to determine which EMDE scenarios generate observable enhancements to the microhalo population.

This paper is organized as follows. In section \ref{back}, we study the evolution of the different components of the universe in our model, including the density, sound speed and equation of state of the $Y$ particles as they transition from being relativistic to nonrelativistic. In section \ref{pert}, the evolution of the density perturbations in the $Y$ particles and dark matter before, during, and after the EMDE is determined, and the suppression of perturbation growth due to the pressure of the $Y$ particles is analyzed. In section~\ref{peakscale}, we present expressions for the wavenumber of the peak scale, for which the matter power spectrum is maximized. In section~\ref{tf}, we provide fitting forms for transfer functions for the computation of the matter power spectrum in scenarios with an initially relativistic particle dominating the hidden sector. Section \ref{peakampl} presents calculations of the dark matter annihilation boost and the power spectrum peak height using our transfer functions; we also discuss prospects for detecting the microhalos generated in EMDE cosmologies. Our results are summarized in section \ref{summ}. The full calculation of the density, pressure, and sound speed of the $Y$ particles is presented in Appendix \ref{ysol-method}. Appendix \ref{impres} contains the derivations of several relations between the parameters that describe the EMDE and the properties of the $Y$ particle. The equations that govern the evolution of perturbations and their initial conditions are detailed in Appendix \ref{pertsol}. Finally, we provide an online application for computing EMDE-enhanced power spectra with the accurate small-scale cut-off that is described in Appendix \ref{app}. This paper uses natural units throughout, in which $c = \hbar = k_{\mathrm{B}} = 1$.

\section{Evolution Of The Homogeneous Background}
\label{back}

Our model considers a universe with three components: dark matter $X$; the thermal bath of relativistic SM particles, which we call radiation (denoted by the subscript $R$); and a particle $Y$ with mass $m$ that decays into SM particles. 
$X$ and $Y$ live in a hidden sector that is thermally decoupled from the Standard Model and has its own temperature $T_{\rm hs}$. The $Y$ particles are initially relativistic but transition to nonrelativistic behavior as the temperature of the hidden sector decreases. We assume that the $X$ particles have frozen out before our calculations begin and are nonrelativistic with $m_X \gg T_{\rm hs}$ and $\rho_X(a) \propto a^{-3}$. 

We first establish the evolution of the homogeneous energy densities of the various components of our model. We begin our calculations at scale factor $a_i$, which is chosen such that $  T_{\rm hs,i} \equiv T_{\rm hs}(a_i) = 300m$, so that the $Y$ particles are initially relativistic. The initial SM density is set by the parameter $\eta \equiv \rho_R(a_i) / \rho_Y(a_i)$. The $Y$ particles are weakly coupled to the SM particles with a decay rate $\Gamma$. Such couplings of the hidden sector to the Standard Model can arise via various renormalizable interactions, including the lepton portal \cite{hid7,lep1}, the Higgs portal \cite{higgs1,higgs2,hid7}, and the vector portal \cite{vec1,hid7}. To obtain the evolution of the energy densities of these three components, the coupled equations for $\rho_X$, $\rho_Y$ and $\rho_R$ are solved numerically:\begin{subequations} \label{eqback}
\begin{align}
    \dot{\rho}_Y + 3H(1 + w_Y)\rho_Y &= - \Gamma mn_Y; \label{ry} \\ 
    \dot{\rho}_R + 4H\rho_R &=  \Gamma mn_Y; \label{rr} \\ 
    \dot{\rho}_X + 3H\rho_X &= 0, \label{rx}
    \end{align}
\end{subequations} where overdots denote $d/dt$ and $H \equiv \dot{a}/a$. In Eq.~(\ref{ry}), $n_Y$ is the number density of $Y$ particles, and $w_Y$ is their time-varying equation of state parameter, defined as the ratio between pressure and density, \mbox{$w_Y \equiv P_Y / \rho_Y$}. The time evolution of $w_Y$ encodes the transition from relativistic to nonrelativistic behavior for the $Y$ particles, which we solve for exactly; the process is detailed in Appendix \ref{ysol-method}. The terms on the RHS of Eqs.~(\ref{ry}) and (\ref{rr}) depend on $mn_Y$ instead of $\rho_Y$ because the longer lab-frame lifetimes of faster particles compensate for the higher energies released by their decays \cite{cannibal_big}. We assume that $X$ and $Y$ particles are coupled only gravitationally, with no momentum exchange between the two species. However, the effects of additional couplings are discussed in Sec. \ref{coupl}.

\begin{figure}[h!]
\centering
\includegraphics[scale=1]{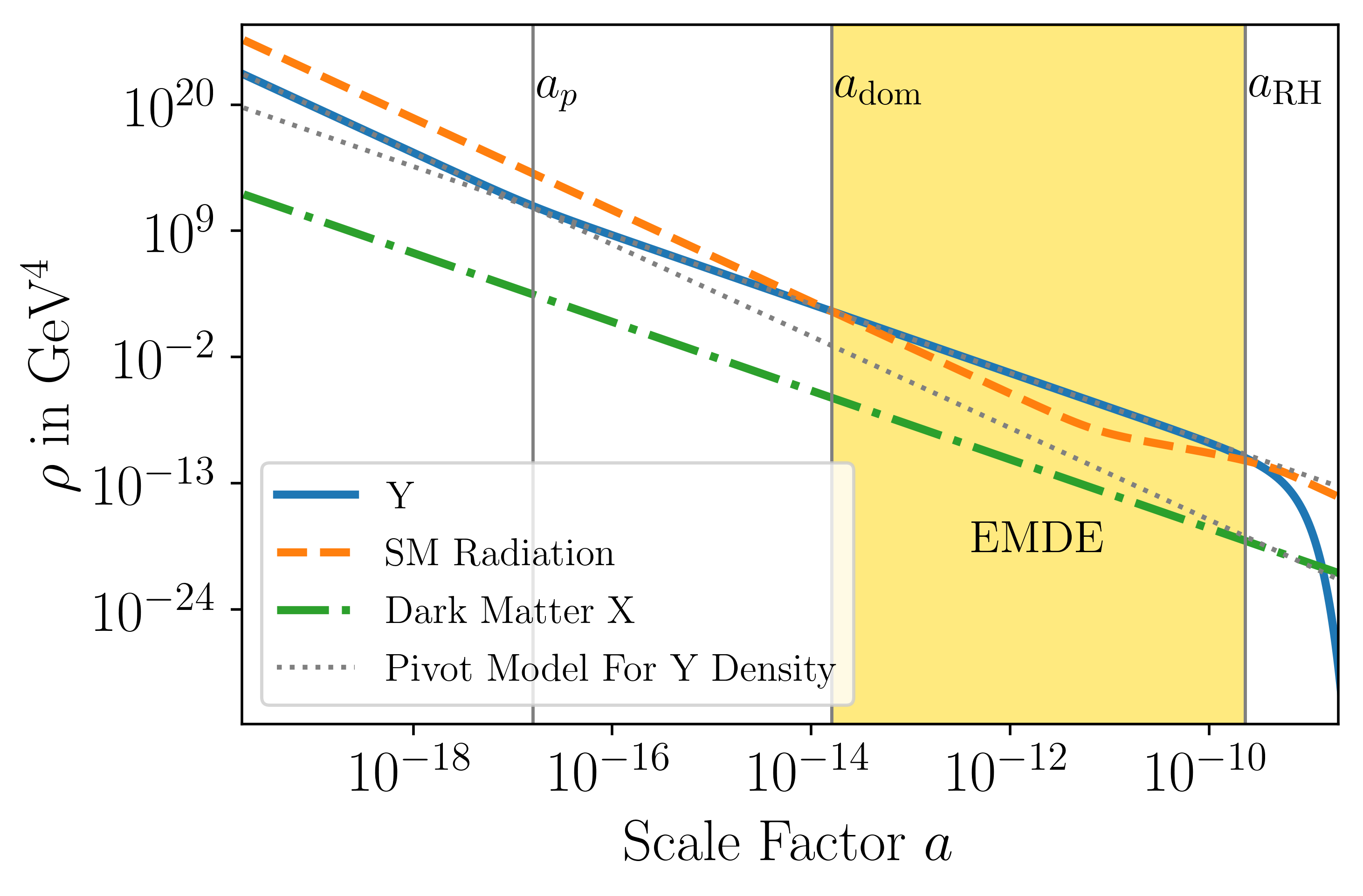}
\caption{The background evolution of the energy densities of the $Y$ particles, SM radiation and dark matter ($X$) as a function of scale factor, for parameters $m =  2$ TeV and $\eta = 1000$.  The pivot scale factor $a_p$ marks the transition from $\rho_Y \propto a^{-4}$ to $\rho_Y \propto a^{-3}$ .  The yellow shaded region shows the EMDE, which begins at the scale factor $a_{\rm dom}$. At the end of the EMDE, $\rho_Y$ rapidly decreases, and the universe becomes radiation dominated.}
\label{fig-back}
\end{figure}

Figure~\ref{fig-back} shows the solutions to Eqs.~(\ref{eqback}) for a chosen set of parameters. 
The transition from $\rho_Y(a) \propto a^{-4}$ to $\rho_Y \propto a^{-3}$ can be modeled by a broken power law with a pivot scale factor given by $a_p/a_{i} = b T_{\rm hs,i}/m$ where $b$ depends only on the statistics of the $Y$ particles. We find that $b$ is 2.70 for bosons and 3.15 for fermions; these values of $b$ are derived in Appendix \ref{ysol-method}. 
It follows from Eq.~(\ref{rr}) that $\rho_R \propto \gs{T} T^4 \propto a^{-4}$ when $\Gamma mn_Y \ll H \rho_R$, where $\gs{T}$ is the relativistic degrees of freedom contributing to the energy density of relativistic SM particles. However, all our analytical results assume that entropy is conserved in the visible sector when $\Gamma mn_Y  \ll H \rho_R$, so that $g_{*S}(T) a^3 T^3$ is constant, where $g_{*S}$ is the relativistic degrees of freedom contributing to the entropy density of the SM bath. When $\Gamma mn_Y$ exceeds $H \rho_R$, $\rho_R \propto a^{-3/2}$ due to the entropy injection from the decay of the $Y$ particles into the visible sector. After the $Y$ particles decay away, $\rho_R \propto a^{-4}$ again. 

The EMDE, indicated by the yellow shaded region in Figure~\ref{fig-back}, starts when $\rho_Y$ exceeds $\rho_R$ at the scale factor $a_{\rm dom}$. We parameterize this point by the temperature of the SM radiation $T_{\mathrm{dom}}$, so that $\rho_{R}(a_{\rm dom}) = (\pi^2 / 30) g_{*}(T_{\mathrm{dom}}) T_{\mathrm{dom}}^4$. We show in Appendix \ref{impres} that $T_{\rm dom}$ can be expressed in terms of our model parameters as 
\begin{equation}
    \gs{T_{\rm dom}}^{\frac{1}{6}} T_{\rm dom} = (fg)^{\frac{1}{4}} \left( \frac{m}{b} \right) \gs{T_i}^{-\frac{1}{12}} \eta^{- \frac{3}{4}},
\end{equation}where $g$ equals the number of degrees of freedom of the $Y$ particles, $T_i$ is the temperature of the SM radiation at $a_i$, and $f$ is 1 if the $Y$ particles are bosons and $7/8$ if they are fermions. 

The EMDE lasts until $\Gamma/H $ becomes comparable to unity. After this point, the comoving number density of the $Y$ particles starts decreasing rapidly. Shortly thereafter, $\rho_Y$ becomes negligible and the universe transitions to radiation domination. This transition, called \textit{reheating}, is not an instantaneous process, but we find it useful to define a reheating temperature $\rh{T}$ in terms of the decay rate as \begin{equation}\label{gamma}
    \Gamma \equiv \sqrt{\frac{8 \pi G}{3} \frac{\pi^2}{30} \gs{\rh{T}} \rh{T}^4} \,,
\end{equation} which sets $\Gamma$ equal to the Hubble rate in a purely radiation-dominated universe at temperature $\rh{T}$. It is also useful to define $\rh{a}$ as the scale factor at which \begin{equation} \label{rhoy}
    \rho_Y (a_p) a_p^3 \equiv a_{\rm RH}^3 \frac{\pi^2}{30} \gs{\rh{T}} \rh{T}^4 .
\end{equation} 
Note that $\rh{T}$ is the quantity defined in Eq.~(\ref{gamma}) and does not equal  $T(\rh{a})$.   

In our broken-power-law model, $\rho_Y(a_p) a_p^4 = \rho_Y(a_i) a_i^4$. Since the $Y$ particles are relativistic at $a_i$, $\rho_Y(a_i) = fg (\pi^2 / 30) T_{\rm hs,i}^4$; it follows from the definition of $\rh{a}$ that \begin{equation}\label{arhap}
    \frac{\rh{a}}{a_p} =\left[ \frac{fg}{\gs{\rh{T}}} \right]^{\frac{1}{3}} \left[ \frac{(m/b)}{\rh{T}} \right] ^{\frac{4}{3}}.
\end{equation} 
To relate $\rh{a}$ to the scale factor today ($a_0$), we note that there is negligible transfer of entropy from the decay of the $Y$ particles to the SM radiation for $a>5\rh{a}$. We find numerically that $T(5\rh{a}) = 0.204 \rh{T}$ and use entropy conservation from $5 \rh{a}$ to $a_0$ to express \begin{equation}\label{arha0}
    \frac{\rh{a}}{a_0} = \frac{1}{1.02} \left[ \frac{\gss{T_0}}{\gss{0.204 \rh{T}}} \right]^{\frac{1}{3}} \left[ \frac{T_0}{\rh{T}} \right] ,
\end{equation} where $T_0$ is the temperature of radiation in the Universe today.

\section{Evolution Of Perturbations}
\label{pert}

The Einstein equations are perturbed to obtain the equations for the evolution of the density contrast $\delta \equiv  (\rho - \bar{\rho}) / \bar{\rho}$ (where $\bar{\rho}$ denotes homogeneous background density) and the velocity dispersion $\theta \equiv a \partial_i v^i$ for each fluid, where $v^i = dx^i / dt$. We work in the Newtonian gauge, in which the metric is given by \begin{equation} ds^2 = -(1 + 2\psi)dt^2 + a^2(t)(1 + 2\phi)(dx^2 + dy^2 + dz^2). \end{equation} We neglect anisotropic stress and set $\psi=-\phi$. The perturbation equations and initial conditions are provided in Appendix \ref{pertsol}. 

\begin{figure}[h!]
    \centering
    \includegraphics[width=0.96\textwidth]{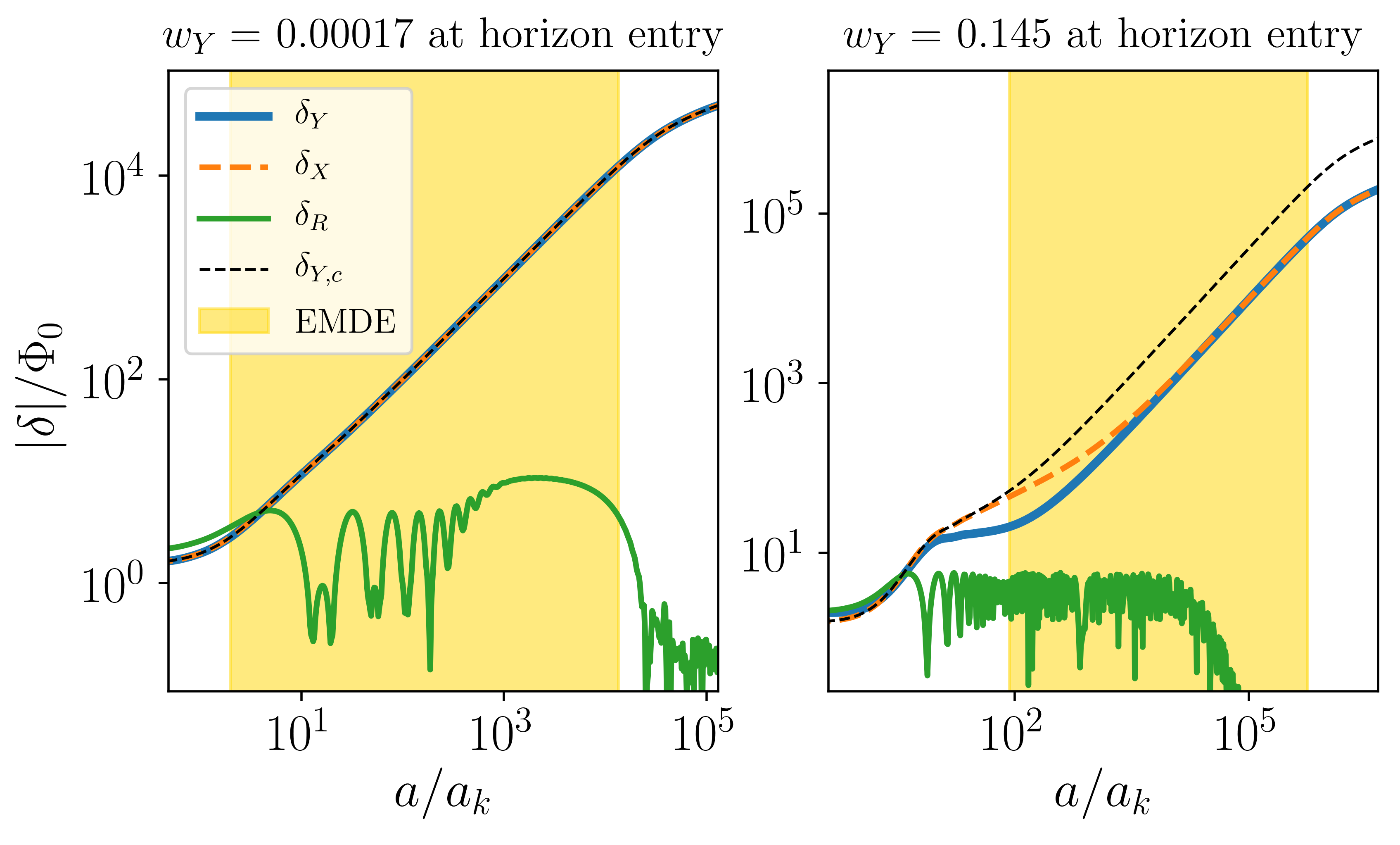} 
    \caption{The evolution of perturbations for two modes, plotted as a function of $a/a_k$, where $a_k$ is the scale factor of horizon entry for the mode. The mode in the left panel enters the horizon after the $Y$ particles have become pressureless ($T_{\rm hs}/m =0.00018$ at horizon entry); there is no suppression of $\delta_Y$ for this mode. The mode in the right panel enters the horizon when the $Y$ particles have significant pressure ($T_{\rm hs} /m = 0.21$ at horizon entry) because of which $\delta_Y$ is suppressed compared to $\delta_{Y,c}$, the $Y$ density perturbation if the $Y$ particles are nonrelativistic.} \label{fig-mode-evol}   
\end{figure}

Figure~\ref{fig-mode-evol} shows the time evolution of $|\delta_i| / \Phi_0$, where $\Phi_0$ is the primordial metric perturbation in a radiation-dominated universe and $i$ denotes the three fluids in our model. Also shown is the evolution of the $Y$ density perturbation if the $Y$ particles were pressureless ($\delta_{Y,c}$). The left panel shows a mode that enters the horizon after the $Y$ particles have become nonrelativistic, with $T_{\rm hs} /m = 0.00018$ at horizon entry. In the absence of pressure, subhorizon density perturbations in $Y$ grow logarithmically with scale factor during radiation domination and linearly during the EMDE. After the EMDE, radiation domination resumes and $\delta_X$ and $\delta_Y$ start growing logarithmically. For this mode, $\delta_Y$ coincides with $\delta_{Y,c}$ because the $Y$ particles are already pressureless when the mode enters the horizon. In contrast, the right panel of Figure~\ref{fig-mode-evol} shows a mode that enters the horizon when the $Y$ particles have significant pressure, with $w_Y = 0.14$ and $T_{\rm hs} /m = 0.21$ at horizon entry. For this mode, the growth of $\delta_Y$ is suppressed compared to that of $\delta_{Y,c}$ until the $Y$ particles become pressureless. As a result, $\delta_Y$ starts linear growth later than $\delta_{Y,c}$ and $\delta_Y < \delta_{Y,c}$ at the end of the EMDE. 

The right panel of Figure \ref{fig-mode-evol} also shows how the evolution of $\delta_X$ is affected by the pressure of the $Y$ particles.  When the mode enters the horizon during radiation domination, $\delta_X$ starts to grow logarithmically with the scale factor.  The pressure of the $Y$ particles delays the onset of linear growth during the EMDE because the $Y$ particles are not as clustered as they would have been if $\delta_Y$ had also grown logarithmically prior to the EMDE.  Instead of growing linearly with scale factor throughout the EMDE, $\delta_X$ converges to $\delta_Y$ because the $X$ particles fall into the gravitational wells generated by the $Y$ particles. Due to this convergence, we will focus hereafter on analyzing the behavior of $\delta_Y$. 

\begin{figure}[h!]
\centering
\includegraphics[scale=0.8]{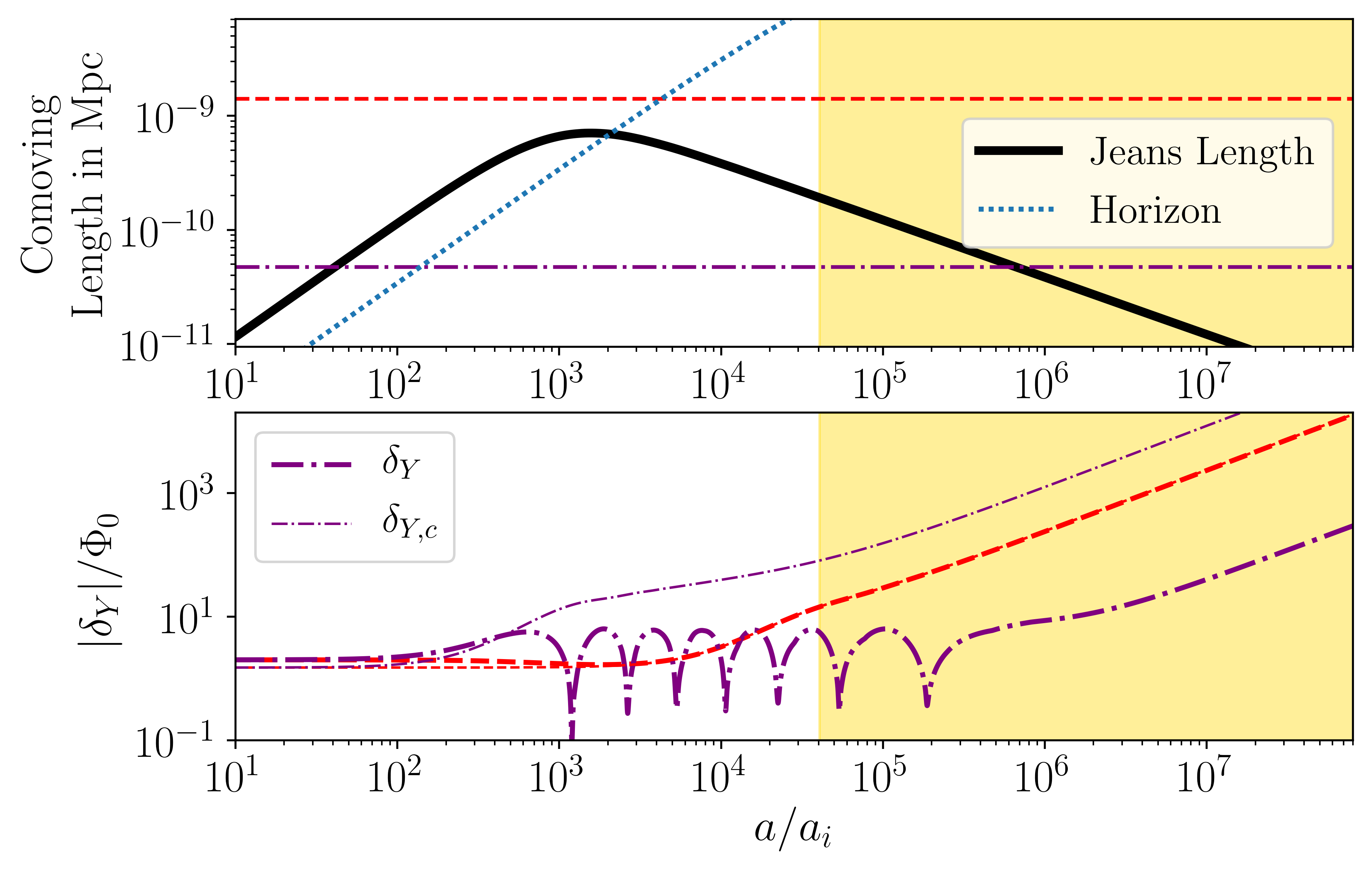}
\caption{\textbf{Top}: The comoving wavelengths of two modes (horizontal lines), placed relative to the comoving Jeans length $\lambda_{\rm J} \equiv k_{\rm J}^{-1}$ and the comoving horizon $(aH)^{-1}$. The yellow shaded region is the EMDE.  \textbf{Bottom}: The thick curves show the evolution of perturbations in the case where $Y$ particles have relativistic pressure and the thin curves show the case where $Y$ particles are treated as cold.  The dashed curve corresponds to a scale that is always larger than the Jeans length, while the dot-dashed curve shows a scale that is much smaller than the Jeans length when it enters the horizon.}
\label{fig-jeans}
\end{figure}

To quantify which scales undergo growth suppression, we consider the continuity and Euler equations for the evolution of density and velocity perturbations in the $Y$ particles along with the Poisson equation. Since the comoving number density of $Y$ particles remains constant until $\Gamma \simeq H$ at the end of the EMDE, we can neglect the decay terms when the pressure of the $Y$ particles is significant. We then have the following equations (taken from Appendix \ref{pertsol}): \begin{subequations} \label{yr-back}
\begin{align}
    \frac{d\delta_Y}{da} =& -(1 + w_Y)\left(\frac{\theta_Y}{a^2 H} + 3\frac{d\phi}{da}\right) - \frac{3}{a} (c_{sY}^2 - w_Y)\delta_Y , \label{dyeq} \\ 
    \frac{d\theta_Y}{da} =& - \frac{1}{a} (1 - 3w_Y) \theta_Y - \frac{dw_Y}{da}\frac{\theta_Y}{1+w_Y} + \frac{c_{sY}^2 k^2 \delta_Y}{(1 + w_Y) a^2 H} - \frac{k^2 \phi}{a^2 H}, \label{thyeq} \\ 
    a \frac{d\phi}{da} =& -\left(1+\frac{k^2}{3a^2H^2}\right) \phi +  \frac{4 \pi G} {3H^2} (\rho_Y \delta_Y + \rho_R \delta_R), \label{poiss}
    \end{align} 
\end{subequations} where $c_{sY}^2 = \delta P_Y / \delta \rho_Y$ is the sound speed of the $Y$ particles (see Appendix \ref{ysol-method}). In Eq.~(\ref{poiss}), the contribution of the dark matter term ($\delta_X \rho_X$) on the RHS is neglected because $\rho_X \ll \rho_Y$. Working in the subhorizon limit where $k \gg aH$ and using $H^2 = (8\pi G/3) (\rho_Y + \rho_R)$, Eq.~(\ref{poiss}) implies that
 \begin{equation} \label{poiss3}
    \left(\frac{k}{aH} \right)^2 \phi \simeq \frac{3}{2} \frac{\rho_Y \delta_Y + \rho_R \delta_R}{\rho_Y + \rho_R}.
\end{equation} 
To obtain the evolution of $\delta_Y$, we neglect the derivative of $\phi$ in Eq.~(\ref{dyeq}) as it is small compared to $\theta_Y / (a^2H)$ and neglect the $(c_{sY}^2 - w_Y) \delta_Y$ term in Eq.~(\ref{dyeq}) since $c_{sY}^2 - w_Y \approx 0$. Similarly, the term proportional to $(dw_Y/da) \theta_Y$ in Eq.~(\ref{thyeq}) is neglected because $dw_Y/da$ is of the order of $(c_{sY}^2 - w_Y)$. Differentiating Eq.~(\ref{dyeq}) with respect to $a$ and using Eqs.~(\ref{thyeq}) and (\ref{poiss3}) gives \begin{equation}\label{2ode-dy}
    \frac{d^2 \delta_Y}{da^2}+ \frac{1}{a}\left[ \frac{d{(a^2 H)}/da}{a H} + (1 - 3w_Y) \right] \frac{d\delta_Y}{da} + \frac{1}{a^2}\left[ \frac{c_{sY}^2 k^2}{a^2 H^2} -  \frac{3}{2} \frac{(1 + w_Y)\rho_Y}{\rho_R + \rho_Y} \right] \delta_Y = \frac{3}{2a^2} \frac{(1 + w_Y)\delta_R \rho_R}{\rho_R + \rho_Y}.
\end{equation} As Figure~\ref{fig-mode-evol} shows, $\delta_R$ begins oscillating shortly after the mode enters the horizon. The gravitational contribution of the $\delta_R \rho_R$ term on the RHS of Eq.~(\ref{2ode-dy}) thus averages to zero and the term can be ignored. We can then express Eq.~(\ref{2ode-dy}) as \begin{equation}\label{eqjeans}
    \frac{d^2 \delta_Y}{da^2}+ \frac{1}{a}\left[ \frac{d{(a^2 H)}/da}{a H} + (1 - 3w_Y) \right] \frac{d\delta_Y}{da} + \frac{c_{sY}^2}{(a^2 H)^2}\left[k^2 - k_{\rm J}^2 \right] \delta_Y = 0,
\end{equation} where we define the time-varying Jeans wavenumber \begin{equation}
    k_{\rm J}^2 (a) \equiv \frac{3}{2} \frac{1 + w_Y}{c_{sY}^2} \frac{\rho_Y}{\rho_R + \rho_Y} a^2 H^2.
\end{equation} When the $Y$ particles are relativistic, $k_{\rm J}$ is roughly proportional to $a^{-1}$ because $c_{sY}^2$ is constant and $\rho_Y \propto a^{-4}$. As the $Y$ particles become colder, $k_{\rm J}$ increases proportional to $a^{1/2}$ because $c_{sY}^2 \propto a^{-2}$ and $\rho_Y \propto a^{-3}$. This behavior is shown in the top panel of Figure~\ref{fig-jeans}, where the black line shows the Jeans length $\lambda_{\rm J} \equiv k_{\rm J}^{-1}$.  

The sign of the coefficient of $\delta_Y$ in Eq.~(\ref{eqjeans}) determines whether $\delta_Y$ grows or oscillates. Figure~\ref{fig-jeans} illustrates the contrast between the growing and oscillating solutions. The top panel shows the comoving length scales ($k^{-1}$) corresponding to two different modes, plotted relative to the Jeans length. The bottom panel shows the time evolution of $\delta_Y$ for the two modes. The thin lines show the evolution of each mode if the $Y$ particles are treated as nonrelativistic ($\delta_{Y,c}$). When $k < k_{\rm J}$ (so that $k^{-1} > \lambda_{\rm J}$), the coefficient of $\delta_Y$ in Eq.~(\ref{eqjeans}) is negative, which leads to a growing solution for $\delta_Y$. The mode represented by the red dashed line in Figure~\ref{fig-jeans} is such an example; its wavelength is always larger than the Jeans length. The bottom panel shows how the amplitude for this mode grows logarithmically with $a$ during radiation domination and then grows linearly with $a$ during the EMDE. In contrast, for the mode indicated by the purple dot-dashed line, the perturbation amplitude oscillates when $k^{-1}<\lambda_J$ and starts growing when $k^{-1}>\lambda_J$. Since $\delta_Y$ starts growing only when the Jeans length becomes smaller than the mode wavelength, $\delta_Y$ is reduced compared to $\delta_{Y,c}$.

\begin{figure}[h!]
\centering
\includegraphics[scale=1]{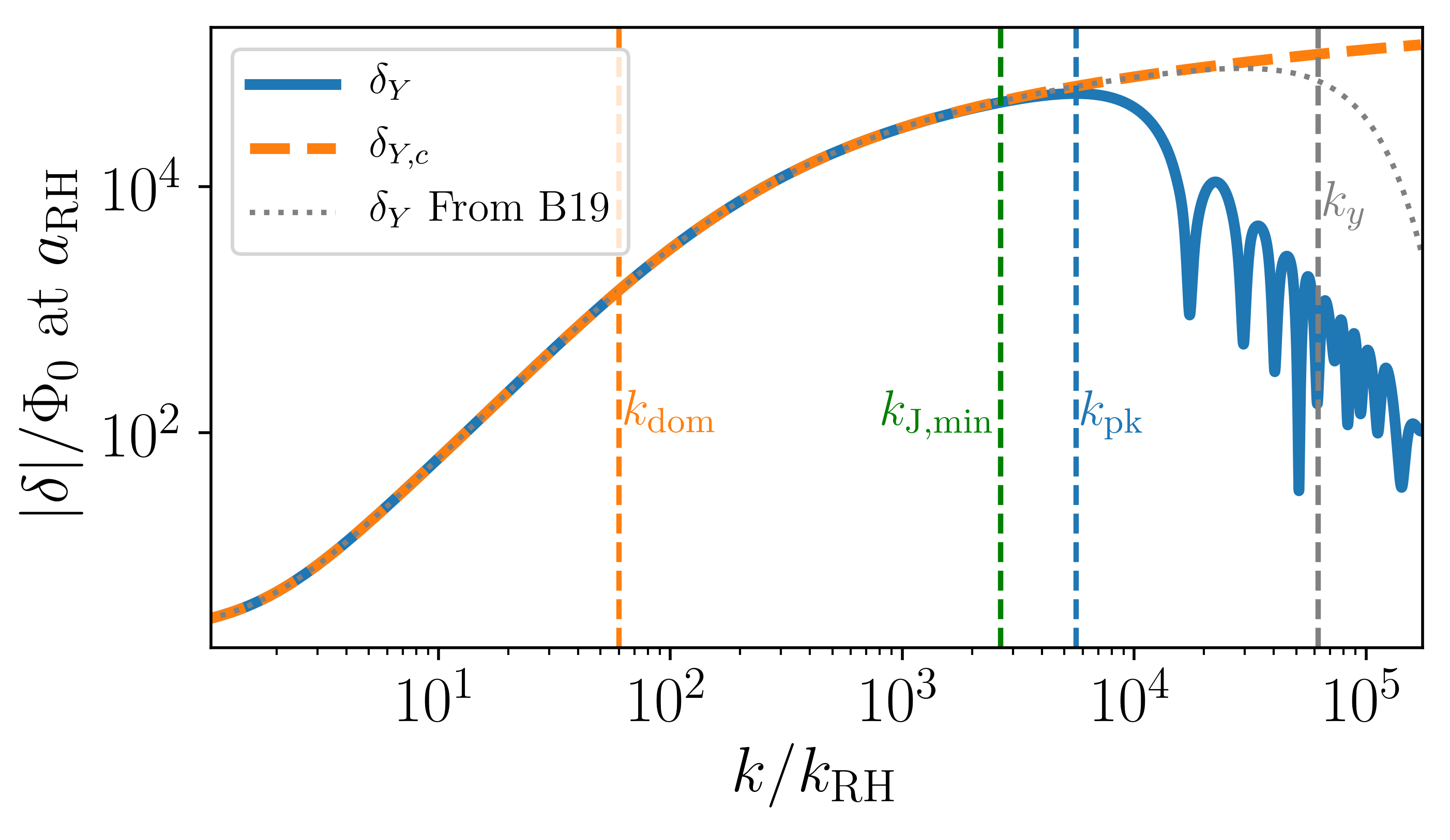}
\caption{Density perturbations in $Y$ at $\rh{a}$ as a function of wavenumber, with $m = 5$ TeV, $\rh{T} = 20$ MeV and $\eta = 500$. The blue curve shows the solution in the case where the $Y$ particles are initially relativistic ($\delta_Y$). The orange dashed curve shows the case when they are treated as pressureless for the same $\rh{T}$ ($\delta_{Y,c}$). The dot-dashed grey curve shows the estimate of $\delta_Y(k)$ used by B19: $\delta_{Y,c}(k)\exp[-k^2/(2k_y^2)]$, where $k_y$ is the horizon wavenumber when $m = T_{\rm hs}$.}
\label{fig-deltayk}
\end{figure}

The suppression of perturbation modes that enter the Jeans horizon is readily apparent in Figure \ref{fig-deltayk}, which shows $\delta_Y$ and $\delta_{Y,c}$ evaluated at $\rh{a}$ as a function of wavenumber scaled by $\rh{k}\equiv \rh{a} \Gamma$. If the $Y$ particles are always pressureless, modes that enter the horizon before the EMDE grow logarithmically with $a$ during radiation domination and then linearly with $a$ during the EMDE, so that $\delta_{Y,c}(k>k_{\rm dom},\rh{a}) \propto \ln (k/k_{\rm dom})$, where $k_{\rm dom} \equiv a_{\rm dom}H(a_{\rm dom})$ is the horizon wavenumber at $a_{\rm dom}$. Modes that enter the horizon during the EMDE grow linearly with $a$ from horizon entry until $\rh{a}$, so that $\delta_{Y,c}(\rh{k} < k < k_{\rm dom},\rh{a}) \propto (k/\rh{k})^2$. The shape of $\delta_{Y,c}(k)$ after the EMDE only depends on the ratio $k_{\rm dom}/ \rh{k}$, which (as shown in Appendix \ref{impres}) can be expressed in terms of our model parameters as 
\begin{equation}\label{kd-krh}
    \frac{k_{\mathrm{dom}}}{\rh{k}} = \sqrt{2} \left[ \frac{gf}{\gs{\rh{T}}} \right]^{\frac{1}{6}} \left[ \frac{\gs{T_i}}{\gs{T_{\rm dom}}} \right]^{\frac{1}{6}}\left[ \frac{(m/b)}{\rh{T}} \right]^{\frac{2}{3}}  \eta^{-\frac{1}{2}}.
\end{equation} 

If the $Y$ particles are initially relativistic, the growth of perturbations is suppressed for scales close to or smaller than the maximum value of the Jeans length (shown by the wavenumber $k_{\rm J,min} = \lambda^{-1}_{\rm J,max} $ in Figure \ref{fig-deltayk}). For these modes, $\delta_Y$ does not begin to grow until the Jeans length becomes smaller than the mode's wavelength. As a result, $\delta_Y$ at $\rh{a}$ is increasingly suppressed compared to $\delta_{Y,c}$ as $k$ increases, as the blue curve in Figure~\ref{fig-deltayk} shows. The suppression leads to a peak in $\delta_Y(k)$ at the wavenumber $k_{\rm pk}$. For $k>k_{\mathrm{pk}}$, modes start growing not only later, but also at different points in the oscillation cycles of their amplitudes. This leads to an oscillation pattern in $\delta_Y(k)$ with a decaying envelope. 

B19 modeled the suppression of modes that enter the horizon when the $Y$ particle is relativistic by multiplying $\delta_{Y,c}(k)$ by $\exp[-k^2/(2k_y^2)]$, where $k_y$ is the wavenumber of the mode that enters the horizon when $m=T_{\rm hs}$. In Appendix \ref{impres}, we derive expressions for $k_y/k_{\rm dom}$ for a universe with $\eta>1$. Using the expression for $k_y / k_{\rm dom}$ from Eq.~(\ref{kykd}) with Eq.~(\ref{kd-krh}) yields \begin{equation}\label{kykrh}
    \frac{k_y}{\rh{k}} = b   \left[ \frac{gf}{\gs{\rh{T}}} \right]^{\frac{1}{6}} \left[ \frac{g_{*}^3 (T_i)}{g_{*y} g_{*}^2(T_{\rm dom})} \right]^{\frac{1}{6}} \left[ \frac{(m/b)}{\rh{T}} \right]^{\frac{2}{3}}  (1 + \eta)^{\frac{1}{2}}\,,
\end{equation} where $g_{*y} = \gs{T(a_y)}$, with $a_y/a_i = T_{\rm hs,i}/m$. The cut-off used by B19 does not describe $\delta_Y(k)$ accurately: Figure \ref{fig-deltayk} shows that $\delta_Y$ falls off at smaller wavenumbers than $k_y$. In section~\ref{peakscale}, we derive the model dependence of the actual peak and cut-off scales of $\delta_Y(k)$.

\section{The Peak Scale}
\label{peakscale}

In order to determine the observational signatures of an EMDE, it is necessary to evaluate the location and amplitude of the peak in the matter power spectrum, since this peak sets the masses, formation times, and central densities of the first microhalos \cite{delos2019:firsthalos,obs_blinov}. In this section, we provide expressions for the peak wavenumber $k_{\rm pk}$ for which $\delta_Y(k)/\Phi_0$ is maximized. 

\begin{figure}[!htb]
\centering
\includegraphics[width=\textwidth]{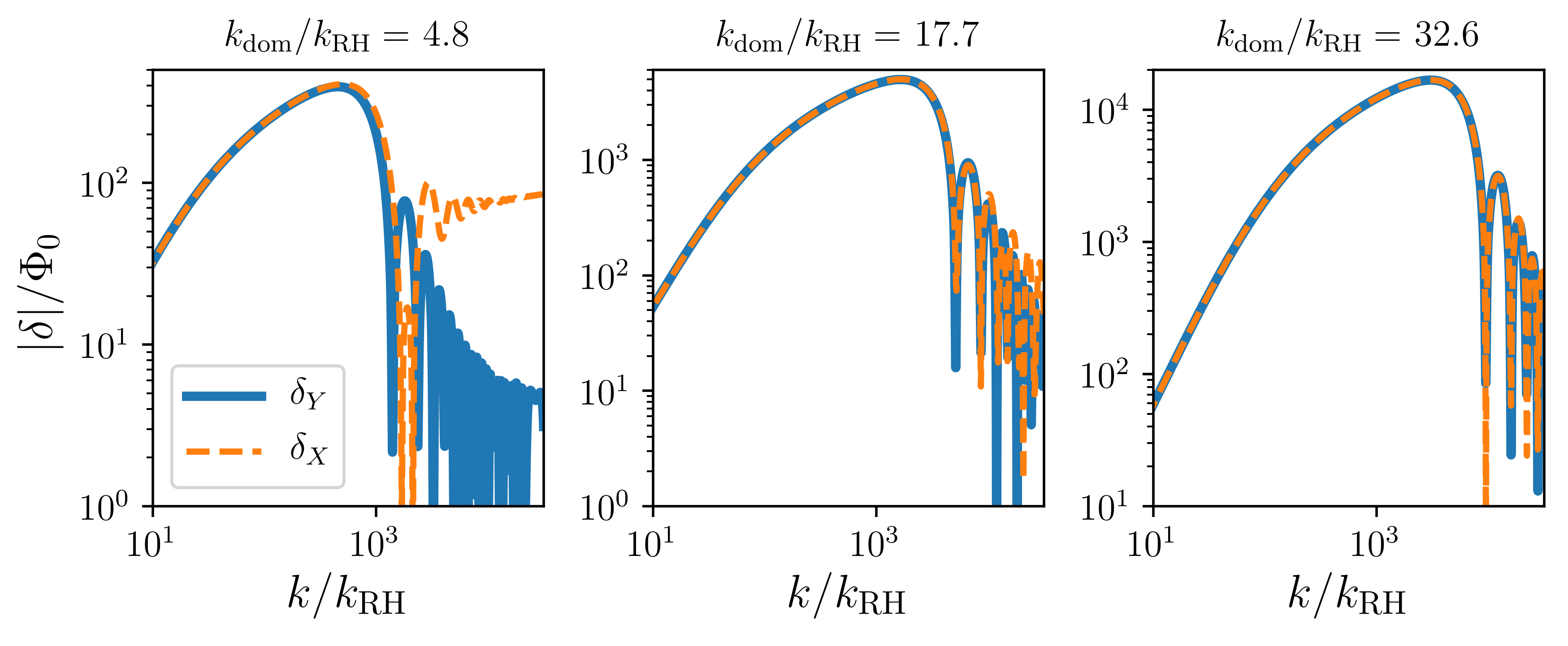}
\caption{Density perturbations at $\rh{a}$ as a function of wavenumber for three different EMDE durations. }
\label{fig-xy}
\end{figure}

Due to the gravitational coupling between $X$ and $Y$ particles during the EMDE, the peak wavenumber of $\delta_X(k)$ is generally very close to that of $\delta_Y(k)$. However, the peaks are not exactly equal in all cases. The relative closeness of the peaks of $\delta_Y(k,\rh{a})$ and $\delta_X(k,\rh{a})$ depends on the duration of the EMDE, quantified by $k_{\rm dom} / \rh{k}$. Figure \ref{fig-xy} shows $\delta_X (k,\rh{a})$ and $\delta_Y(k,\rh{a})$ for three different EMDE durations. The leftmost panel shows a short EMDE with $k_{\rm dom} / \rh{k} = 4.8$, in which case the peak wavenumbers of $\delta_Y$ and $\delta_X$ differ by 10\% at the end of the EMDE. This difference arises because the EMDE is too short for $\delta_X$ and $\delta_Y$ to become equal for modes close to the peak wavenumbers.  For scales smaller than the second peak in the left panel of Figure \ref{fig-xy}, $\delta_Y$ oscillates throughout the EMDE because the Jeans length does not fall below the comoving wavelengths of these modes before $\rh{a}$. As a result, the $Y$ particles do not cluster and never exert a coherent gravitational pull on the $X$ particles. The $X$ particles drift during the EMDE and $\delta_X$ does not approach $\delta_Y$. For longer EMDEs, the peaks of $\delta_X$ and $\delta_Y$ are nearly identical. The middle panel of Figure \ref{fig-xy} shows the case with $k_{\rm dom} / \rh{k} = 17.7$, for which the peak wavenumbers of $\delta_X$ and $\delta_Y$ differ by 1.3\%. For $k_{\rm dom} / \rh{k} = 32.6$, this discrepancy between the peak scales falls to 0.4\%.   Therefore, the peak in $\delta_Y$ generally matches the corresponding peak in the matter power spectrum after the EMDE.

\subsection{The Effect Of Kinetic Coupling In The Hidden Sector}
\label{coupl}

Thus far, we have assumed that the $Y$ particles and the dark matter $X$ are coupled only gravitationally. In this section, we explore how scatterings between $X$ and $Y$ particles affect the peak amplitude and scale of $\delta_X$ and $\delta_Y$.

If the $X$ and $Y$ particles are initially kept in kinetic equilibrium through a scattering process, the momentum transfer rate $(dp/dt)/p$ to the $X$ particles from this scattering is given by $n_Y (m/m_X) \langle\sigma v\rangle $, where $\langle\sigma v\rangle$ is the velocity-averaged scattering cross section. This interaction modifies the Euler equation for the velocity perturbations in the dark matter \cite{dmbar}:
\begin{equation} \label{thxc}
    \theta_X ' = -  \theta_X  - \frac{k^2}{aH} \phi + n_Y \frac{m}{m_X} \frac{\langle \sigma v \rangle}{H} (\theta_Y - \theta_X),
\end{equation} where the prime denotes $d / d\ln a$. 
The corresponding coupling term in the Euler equation for $\theta_Y$ is suppressed by a factor of $\rho_X / \rho_Y$  and can be neglected.
The coupling strength is parameterized by the scale factor of kinetic decoupling $a_{kd}$, which is defined by the relation $n_Y(a_{kd})\langle \sigma v \rangle = H(a_{kd})$.

To study the effect of this kinetic coupling, we consider three examples with $\eta=300$, \mbox{$k_{\rm dom}/\rh{k} = 36$}, and \begin{enumerate}
\itemsep0em 
    \item no kinetic coupling between $X$ and $Y$ particles,
    \item kinetic coupling with $a_{kd} = 0.5 a_{\rm dom} = 150 a_p$, such that the $X$ and $Y$ particles decouple before the EMDE starts but after the $Y$ particles have become cold, and
    \item kinetic coupling with $a_{kd} = 1.4 \rh{a}$, such that the $X$ and $Y$ particles remain coupled until after the $Y$ particles have decayed into SM radiation.
\end{enumerate} While $n_Y \langle \sigma v \rangle \gg H$, $\theta_X \simeq \theta_Y$. As the hidden sector temperature decreases, the $Y$ particles become nonrelativistic. In this regime, we can use the results of Ref. \cite{dmbar} for the momentum transfer rate for the collision of two nonrelativistic particles and take $\langle \sigma v \rangle \propto \sqrt{T_{\rm hs}}$. For all these cases, we choose $m_X/m = 50$ so we can assume the DM particle is much heavier than the $Y$ particle and $w_X \approx 0$. This ensures that the evolution of $\rho_X$ is given by Eq.~(\ref{rx}). 

\begin{figure}[!htb]
\centering
\includegraphics[width=\textwidth]{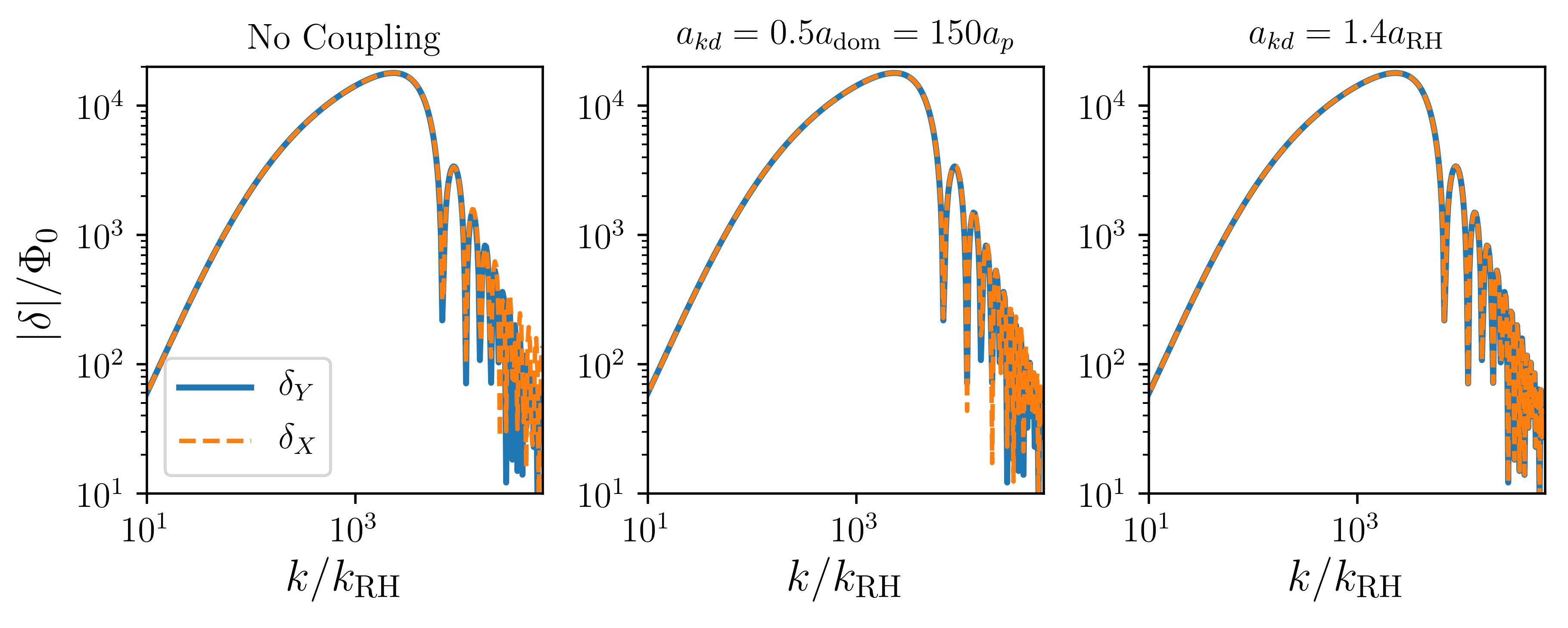}
\caption{Density perturbations as a function of wavenumber, evaluated at $\rh{a}$, with $\eta = 300$, $k_{\rm dom}/\rh{k} = 36$ and $m_X/m = 50$. $\delta_X(k,\rh{a})$ is plotted for cases with three different kinetic coupling strengths. The labels mark the scale factor of kinetic decoupling $a_{kd}$ for the different coupling scenarios. For the case with no coupling (leftmost panel), the peak wavenumbers of $\delta_X$ and $\delta_Y$ differ by 0.4\%. }
\label{fig-coupling}
\end{figure}
 
Figure~\ref{fig-coupling} shows $\delta_Y(k,\rh{a})$ and $\delta_X(k,\rh{a})$ for the three cases mentioned above. The amplitude and the location of the first peak remain the same between the cases. Gravitational coupling during the EMDE is strong enough to make $\delta_X$ and $\delta_Y$ converge to within a 0.5\% difference for wavenumbers close to $k_{\rm pk}$, even without kinetic coupling.  The left panel of Figure~\ref{fig-coupling} also shows that $\delta_Y$ and $\delta_X$ differ for scales smaller than the peak scale in the case with no kinetic coupling. For these modes, $\delta_Y$ oscillates for a portion the EMDE, and the $Y$ particles do not exert a coherent gravitational force on the DM until $\delta_Y$ stops oscillating. As a result, $\delta_X$ has not fully converged to $\delta_Y$ at $\rh{a}$.
Comparing the left and middle panels of Figure \ref{fig-coupling}, it is apparent that scatterings tie $\delta_X$ to $\delta_Y$ for these modes.  Therefore non-gravitational interactions between the $X$ and $Y$ particles only serve to tighten the correspondence between $\delta_Y$ and $\delta_X$ and do not significantly affect the matter power spectrum.

\subsection{Scenarios With Initially Subdominant $Y$ particles}
\label{pk-rd}

To derive an analytical expression for $k_{\rm pk}$ when the $Y$ particle is initially subdominant ($\eta > 1$), we adopt the approach used to find the peak scale for cannibalistic hidden sector particles \cite{cannibal_big}.
The peak scale enters the horizon after the $Y$ particles have become nonrelativistic but before the onset of the EMDE. Since $\delta_R$ oscillates rapidly after the peak scale enters the horizon, the term proportional to $\delta_R \rho_R$ in Eq.~(\ref{2ode-dy}) does not affect the evolution of $\delta_Y$. In addition, the term proportional to $\rho_Y/(\rho_Y + \rho_R)$ in the coefficient of $\delta_Y$ on the LHS is negligible since $\rho_Y \ll \rho_R$ prior to the EMDE. Using entropy conservation in the visible sector, we can write $a^2 H = a_i^2 H(a_i) [\gs{T(a_i)}/\gs{T(a)}]^{1/6}$. The first term in the coefficient of $d \delta_Y / da$ in Eq.~(\ref{2ode-dy}) is then proportional to $d \ln \gs{T(a)} / d \ln a$, which is negligible. We also set $1 + w_Y$ to unity, since $w_Y$ is small compared to 1 and decreases as $a^{-2}$ when the modes close to the peak scale enter the horizon. With these approximations, Eq.~(\ref{2ode-dy}) can be written as \begin{equation} \label{dy-osc}
    \frac{d^2 \delta_Y}{da^2} + \frac{1}{a} \frac{d \delta_Y}{da} + \frac{c_{sY}^2 k^2 \delta_Y}{(a^2H)^2}  = 0.
\end{equation} 

In Appendix \ref{ysol-method}, we present a piecewise model for $c_{sY}^2 (a)$: $c_{sY}^2 (a) = 0.33 a_{pc}^2  /a^{2}$ for $a \gtrsim 1.4 a_p$. Here, $a_{pc} = 1.43a_p$ for bosons and $1.41a_p$ for fermions, where $a_p = bT_{\rm hs,i} a_i /m$ is the pivot scale factor for broken power law that models $\rho_Y(a)$ and $b$ is 2.70 for bosonic $Y$ particles and 3.15 for fermionic $Y$ particles. The different factors 1.41 and 1.43 reflect that the $Y$ particles have slightly different pressure for the same value of $a/a_p$ if their statistics are different, which leads to the pivot points for their sound speed being at slightly different values of $a/a_p$. Since the peak scale enters the horizon when $a \gtrsim 1.4a_p$, we can use $c_{sY}^2 (a) = 0.33 a_{pc}^2  /a^{2}$ in Eq.~(\ref{dy-osc}), which then describes a simple harmonic oscillator in $\ln a$ with the $k$-dependent frequency $\omega_k = \beta a_p k g_a^{1/6} / (a_i^2 H_i g_i^{1/6})$, where $g_a = g_{*}(T(a))$, $g_i = g_{*}(T(a_i))$ and $\beta = 0.82$ for bosons and $\beta=0.80$ for fermions. The factor of $g_a^{1/6}$ introduces a slight time-dependence into $\omega_k$; we neglect this variation and set $g_a = g_k \equiv g_*(T(a_k))$ when solving Eq.~(\ref{dyka}). The solution is \begin{equation} \label{dyka}
    \delta_Y(k,a) = A_1 \sin \left[ \omega_k \ln \left(A_2 \frac{a}{a_k}\right) \right].
\end{equation} 
 
Since $\omega_k$ encodes the effect of the relativistic pressure of the $Y$ particles, the expression for $\delta_Y$ for small $\omega_k$ should match the evolution of cold dark matter in radiation domination \cite{hu96}: $\delta_X(a) = A \Phi_0 \ln (B a / a_k) $ with $A = 9.11$ and $B = 0.594$.  The coefficients $A_1$ and $A_2$ are determined by evaluating Eq.~(\ref{dyka}) when $\omega_k \ll 1$ and matching it to this function.  Prior to the EMDE, $\ln(a/ a_{k}) \lesssim 10$ for modes near the peak scale, and the argument within the sine in Eq.~(\ref{dyka}) is small compared to unity if $\omega_k \ll 1$.   Using the approximation that $\sin x \simeq x$ for $x\ll1$, it follows that $A_1 = A \Phi_0 / \omega_k$ and  $A_2 = B $. 

The peak wavenumber $k_{\rm pk}$ can be found by maximizing the amplitude $\delta_Y(k,a_{\rm dom})$. Using the expressions for $A_1$ and $A_2$, we have \begin{equation} \label{dy-adom}
    \delta_Y(k,a_{\rm dom}) = \frac{A \Phi_0}{\omega_k} \sin \left[\omega_k \ln \left(B \frac{a_{\rm dom}}{a_k} \right)\right] = \frac{A \Phi_0}{\omega_k} \sin \left[\omega_k \ln \left(\sqrt{2} B \gkd \frac{k}{k_{\rm dom}} \right)\right], 
\end{equation} where the second equality results from using the expression for $a_k/a_{\rm dom}$ in radiation domination from Eq.~(\ref{kkd}). Neglecting the weak $k$-dependence of $(g_k/g_{\rm dom})^{1/6}$ and $\Phi_0$ while setting the derivative of Eq.~(\ref{dy-adom}) with respect to $k$ equal to zero implies \begin{equation}
    \tan \left[ \omega_{\rm pk}  \ln \left( \frac{\sqrt{2} B k_{\rm pk}}{k_{\rm dom}} \gpkd \right) \right] = \omega_{\rm pk} \left[ 1 + \ln \left(  \frac{ \sqrt{2} B k_{\rm pk}}{k_{\rm dom}} \gpkd \right) \right].
\end{equation} 

Since $k_{\rm pk}$ is an extremum of $\delta_Y(k)$, the tangent function on the LHS is well-described by a Taylor expansion to second order around $k_{\rm pk}$. Using this expansion and solving the resulting equation for $k_{\rm pk}$ yields \begin{equation} \label{kpk-kd}
    \sqrt{2} B\frac{ k_{\rm pk}}{k_{\rm dom}} \gpkd = \left[ \frac{1.5}{r} W \left( \frac{2r}{3} \right) \right]^{-\frac{3}{2}},
\end{equation} where $W$ is the Lambert W-function and \begin{equation} \label{rdef}
    r = \left[ \frac{\sqrt{6} B }{\beta} \frac{a_i}{a_p} \frac{a_i H_i}{k_{\rm dom}} \left( \frac{g_i}{g_{\rm pk}} \right)^{\frac{1}{6}} \right]^{\frac{2}{3}}.
\end{equation} We express $a_i H_i / k_{\rm dom} = (a_{\rm dom} / (\sqrt{2} a_i)) (g_{\rm dom}/g_i)^{1/6}$ (using Eq.~(\ref{kkd})) and use the definition of $a_{\rm dom}/a_p$ from Eq.~(\ref{adap}) to simplify the dependence of $r$ on the model parameters. Substituting $B = 0.594$, we have \begin{equation}
    r = \left[ \frac{0.95}{\beta} \left( \frac{g_i^2}{g_{\rm pk} g_{\rm dom}} \right)^{\frac{1}{6}} \eta \right]^{\frac{2}{3}}.
\end{equation} Finally, we use the expression for $k_y/k_{\rm dom} $ from Eq.~(\ref{kykd}) to eliminate $k_{\rm dom}$ from Eq.~(\ref{kpk-kd}) and obtain \begin{equation} \label{kpk-rd}
    g_{\rm pk}^{\frac{1}{3}}\frac{k_{\rm pk}}{k_y} = (g_y g_{\rm dom})^{\frac{1}{6}}\frac{1.03}{\beta b} \sqrt{\frac{\eta}{1 + \eta}} W^{-3/2} \left[0.77 \left( \frac{g_i^2}{g_{\rm pk} g_{\rm dom}} \right)^{\frac{1}{9}} \eta^{2/3}\right].
\end{equation} In the above expression, $g_{\rm pk} = \gs{T(a_{\rm pk})}$ and $g_{\rm dom} = \gs{T_{\rm dom}}$.  The RHS of this expression includes an additional factor of 1.08 that brings the $k_{\rm pk}$ values into better agreement with those obtained from the numerical solutions of the perturbation equations.

To establish the relation between the peak scale and our model parameters, the peak scale can be rewritten in physical units. Using Eqs.~(\ref{gamma}) and (\ref{arha0}), and substituting $T_0 = 2.726$ K and $\gss{T_0} = 3.91$ , $\rh{k} \equiv \rh{a} \Gamma$ can be expressed as \begin{equation}\label{arhreal}
\frac{\rh{k}}{a_0} =  7.68 \left[ \frac{g_{*}^{1/2}(\rh{T})}{g_{*S}^{1/3}(0.204 \rh{T})}\right] \left[ \frac{\rh{T}}{\mathrm{1 MeV}} \right] \times 10^3 \, \mathrm{Mpc}^{-1}.
\end{equation}Using the definition of $k_y / \rh{k}$ from Eq.~(\ref{kykrh}), the peak wavenumber is \begin{equation} \label{kpk-a0-rd}
    \frac{k_{\rm pk}}{a_0} = \frac{0.9}{\beta} \left[\frac{f g}{b^4} \right]^{\frac{1}{6}}  \left[ \frac{g_{*}(\rh{T})}{g_{*S}(0.2 \rh{T})}\right]^{\frac{1}{3}} \left[ \frac{\rh{T}}{\mathrm{1 MeV}} \right]^{\frac{1}{3}} \left[ \frac{m}{\mathrm{1 GeV}} \right]^{\frac{2}{3}} \frac{\eta^{\frac{1}{2}}}{  W^{\frac{3}{2}} (0.77\eta^{\frac{2}{3}})} \times 10^6 \,\, \mathrm{ Mpc}^{-1},
\end{equation} where we have ignored the variation of $g_{*}$ before the EMDE for simplicity. In Eq.~(\ref{kpk-a0-rd}),  $k_{\rm pk}/a_0$ depends on $\rh{T}$ because the reheat temperature determines when the EMDE ends and thus affects the expansion history of the Universe after the peak scale enters the horizon.

\begin{figure}[!htb]
\centering
\includegraphics[width = \textwidth]{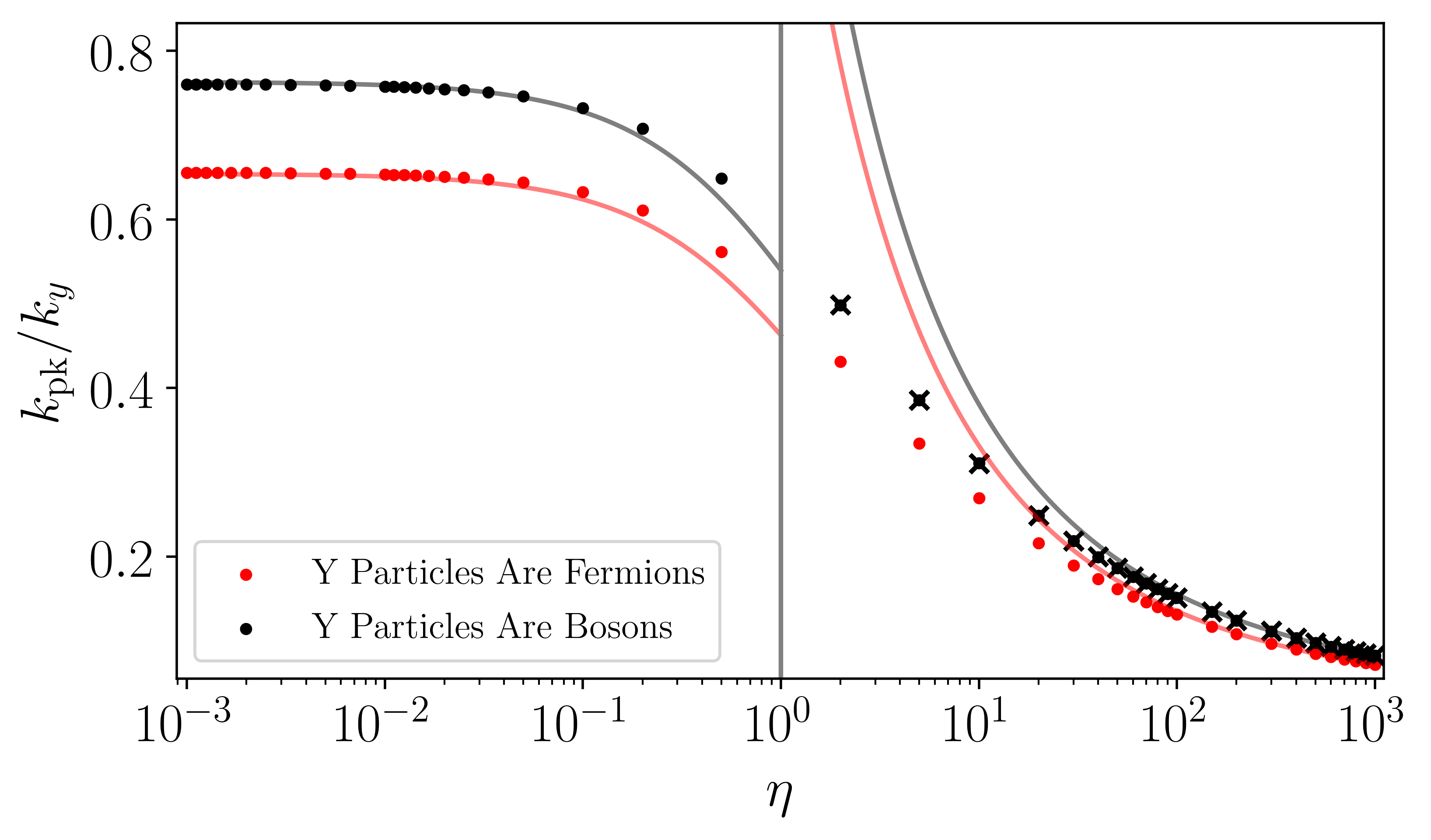}
\caption{The peak wavenumber $k_{\rm pk}$ that maximizes $\delta_Y(k)$, scaled by $k_y$ and plotted against $\eta$, the initial value of $\rho_Y/\rho_R$. The dots represent the $k_{\rm pk}$ determined from numerical data for $m = 2$ TeV and the lines represent the analytical predictions for $k_{\rm pk}/k_y$, given by Eq.~(\ref{kpk-rd}) for $\eta>1$ and by Eq.~(\ref{kpk-yd}) for $\eta<1$. The plotted quantity $k_{\rm pk}/k_y$ depends only on $\eta$ and the statistics of the $Y$ particles. Different colors show cases with different $Y$ particle statistics. This plot assumes $\gs{T} = g_{*S}(T) = 100$. The black crosses show $k_{\rm pk} / k_y$ for a boson $Y$ particle for $m = 200$ GeV; they overlap with the black dots, confirming that $k_{\rm pk} / k_y$ is independent of $m$. }
\label{fig-kpk}
\end{figure}

The points in Figure~\ref{fig-kpk} show $k_{\rm pk}/k_y$ for different $\eta$ values as determined from the numerical solutions for the evolution of $\delta_Y$, while the solid lines for $\eta >1$ show $k_{\rm pk}/k_y$ from Eq.~(\ref{kpk-rd}) with $\gs{T} = 100$.  The analytical expression explains the variation of $k_{\rm pk}/k_y$ with $\eta$ and predicts the peak scale of $\delta_Y(k,\rh{a})$ to within 3\% of the numerically determined peak scale for $\eta \geq 100$.  As $\eta$ decreases, the peak scale enters the horizon closer to the pivot point of $c_{sY}^2$.  Since the asymptotic late-time expression for $c_{sY}^2$ was used in the derivation of Eq.~(\ref{kpk-rd}), its prediction for $k_{\rm pk}$ diverges from the numerically determined peak wavenumber for $\eta<100$.

\subsection{Scenarios With Initial $Y$-Domination}

If $\eta<1$, $\rho_R$ remains subdominant until reheating. The EMDE begins when $\rho_Y$ starts decreasing proportional to $a^{-3}$ at $a=a_p$, and this pivot also determines which modes are suppressed by the relativistic pressure of the $Y$ particles.  The numerical solutions to the perturbation equations for $\eta<0.1$ indicate that $k_{\rm pk}$ enters the horizon while the $Y$ particles are still relativistic ($a_{\rm pk} < a_p$) and that $a_{\rm pk} = a_p\sqrt{1+\eta}/\gamma$, where $\gamma = 2.055$  and $2.065$ for bosonic and fermionic $Y$ particles, respectively.  The factor $\gamma$ accounts for a slight difference between $k_{\rm pk}$ for fermionic and bosonic $Y$ particles, which arises because the $Y$ particles have slightly lower pressure at a given value of $a/a_p$ if they are fermions compared to if they are bosons.  

For $a < a_p$, $H(a) \propto a^{-2}$ and thus $k \propto a_k^{-1}$.    Therefore, $k_{\rm pk} / k_p = \gamma/\sqrt{1 + \eta}$. The wavenumber $k_p \equiv a_p H(a_p)$ can be obtained by expressing $H^2(a_p) = 8 \pi G (\rho_Y(a_p) + \rho_R(a_p)) / 3 = 8 \pi G \rho_Y(a_p)(1 + \eta) / 3 $. In this expression, $\rho_Y(a_p)$ can be written using Eq.~(\ref{rhoy}) and the expression for $a_p / \rh{a}$ from Eq.~(\ref{arhap}). Finally, taking $\Gamma$ from Eq.~(\ref{gamma}), 
 \begin{equation} \label{kpkrh-yd}
    \frac{k_p}{\rh{k}} = \frac{a_p}{\rh{a}} \frac{H(a_p)}{\Gamma} = \left[ \frac{gf}{\gs{\rh{T}}} \right]^{\frac{1}{6}} \left[ \frac{(m/b)}{\rh{T}} \right]^{\frac{2}{3}} \sqrt{1 + \eta}.
\end{equation} 

To express $k_p$ in terms of $k_y$, we again use the scaling $k \propto a_k^{-1}$, which applies since $a_y < a_p$. Using this scaling yields $k_y/k_p = a_p / a_y = b$, so that
\begin{equation} \label{kpk-yd}
     \frac{k_{\rm pk}}{k_y} = \frac{\gamma}{b} \frac{1}{\sqrt{1 + \eta}}.
\end{equation}
This prediction for ${k_{\rm pk}}/{k_y}$ is shown by the solid lines in Figure~\ref{fig-kpk} for $\eta < 1$.
The value given by Eq.~(\ref{kpk-yd}) agrees with the peak scale to within 1\% for $\eta<0.1$. For $\eta>0.1$, $k_{\rm pk}$ enters the horizon after $c_{sY}^2$ begins to decrease. This makes $k_{\rm pk}/k_y$ diverge from the prediction of Eq.~(\ref{kpk-yd}), which is valid for cases in which $c_{sY}^2 (a_{\rm pk}) \simeq 1/3$. We can use Eq.~(\ref{kpkrh-yd}) and $k_y/k_p = b$ in conjunction with the definition of $\rh{k}$ from Eq.~(\ref{arhreal}) to express $k_{\rm pk}$ in physical units as
\begin{equation}
    \frac{k_{\rm pk}}{a_0} = 0.765 \gamma \left[\frac{f g}{b^4} \right]^{\frac{1}{6}}  \left[ \frac{g_{*}(\rh{T})}{g_{*S}(0.2 \rh{T})}\right]^{\frac{1}{3}} \left[ \frac{\rh{T}}{\mathrm{1 MeV}} \right]^{\frac{1}{3}} \left[ \frac{m}{\mathrm{1 GeV}} \right]^{\frac{2}{3}} \times 10^6 \, \, \mathrm{ Mpc}^{-1}.
\end{equation}

\section{Transfer Functions}
\label{tf}

Solving the Boltzmann equations with an initially relativistic $Y$ particle is computationally expensive.  
To facilitate the computation of the matter power spectrum in such hidden-sector cosmologies, we present analytical transfer functions that relate $\delta_Y(k)$ to $\delta_{Y,c}(k)$.  The transfer function is defined as \begin{equation}
	T(k) \equiv \frac{\delta_Y(k)}{\delta_{Y,c}(k)},
\end{equation} where both $\delta_Y(k)$ and $\delta_{Y,c}(k)$ are evaluated at $\rh{a}$. For $\eta>1$, we calculate $\delta_{Y,c}$ by setting $w_Y = c_{sY}^2 = 0$ and $\rho_Y(a_i) = \rho_Y(a_\mathrm{dom})(a_\mathrm{dom}/a_i)^3$, thus obtaining the evolution of density perturbations in cold particles for the same value of $\rho_Y(a)$ during the EMDE.  If the relativistic $Y$ particles initially dominate the universe, making the $Y$ particles cold radically alters the evolution of the Hubble rate.  To avoid conflating the effects of changing the Hubble rate with the effects of the $Y$ particles' pressure, we use an analytical expression for $\delta_{Y,c}$ when computing $T(k)$ when $\eta < 1$, as described in Sec.~\ref{tf-yd}.

\begin{figure}[!htb]
	\centering
	\includegraphics[width=0.8 \textwidth]{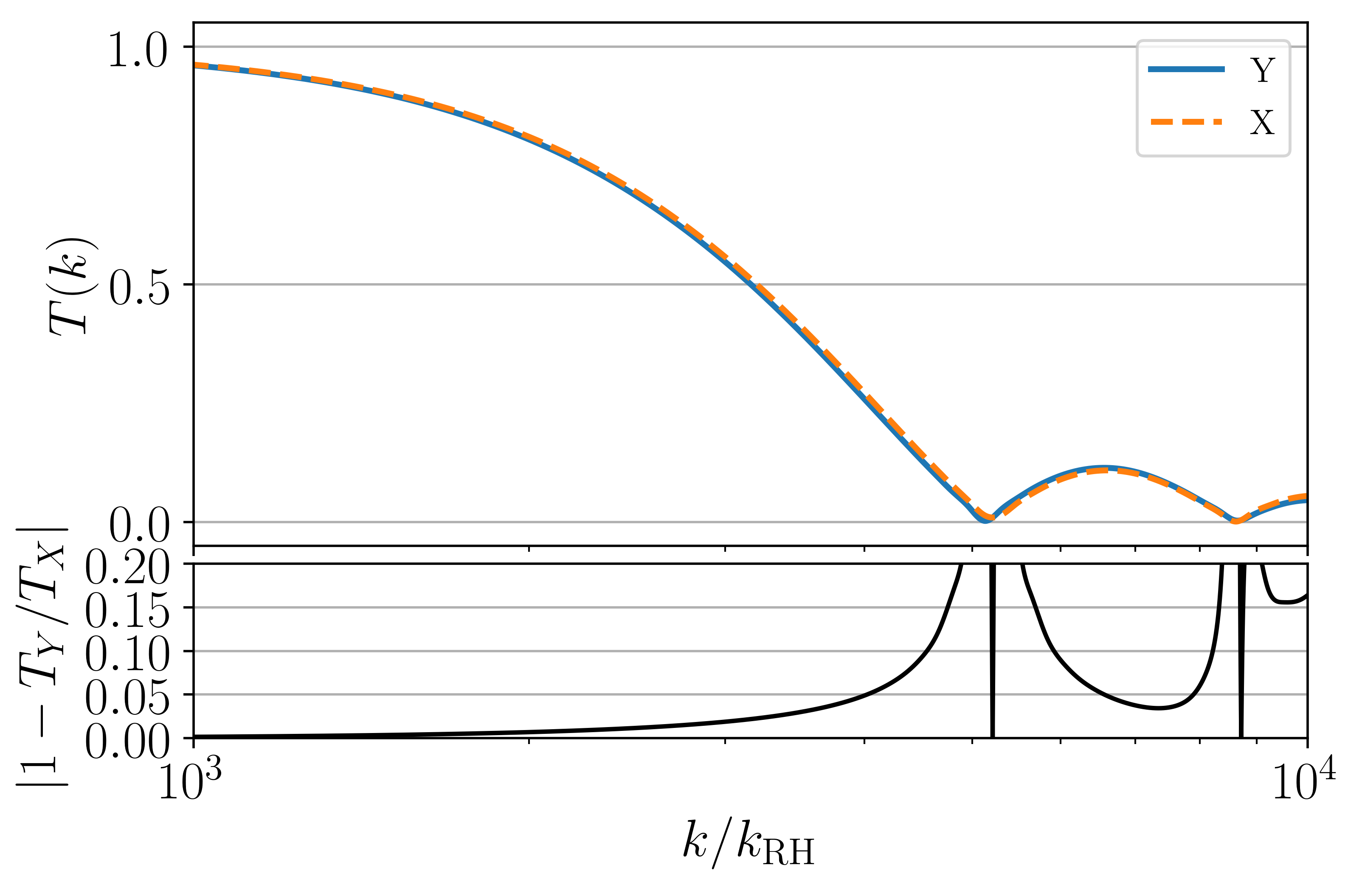}
	\caption{Comparing the transfer functions for $X$ and $Y$ perturbations for a case with $k_{\rm dom} / \rh{k} = 17.7$. In the top panel, the solid line shows $T_Y = T(k)$ and the dashed line shows $T_X(k)$, while the bottom panel shows the relative difference between the two. The two transfer functions agree to within 5\% for $T(k) \geq 0.25$.}
	\label{fig-txy}
\end{figure}

We focus on transfer functions for the $Y$ density perturbations because they are less sensitive to the duration of the EMDE than transfer functions for DM perturbations would be.  In most cases though, these transfer functions can be applied directly to the DM power spectrum.  Figure \ref{fig-txy} shows the correspondence of $T(k)$ and $T_X(k) \equiv \delta_X(k, a_\mathrm{RH}) / \delta_{X,c} (k,a_\mathrm{RH} )$ for $k_{\rm dom} / \rh{k} = 17.7$ (where $\delta_{X,c}$ is the DM perturbation if the $Y$ particles were pressureless). The bottom panel shows the relative error between $T$ and $T_X$, which remains within 5\% for $T(k) \geq 0.25$. Longer EMDEs lead to even closer agreement between $T_X(k)$ and $T(k)$.

\begin{figure}[!htb]
	\centering
	\includegraphics[width=0.95 \textwidth]{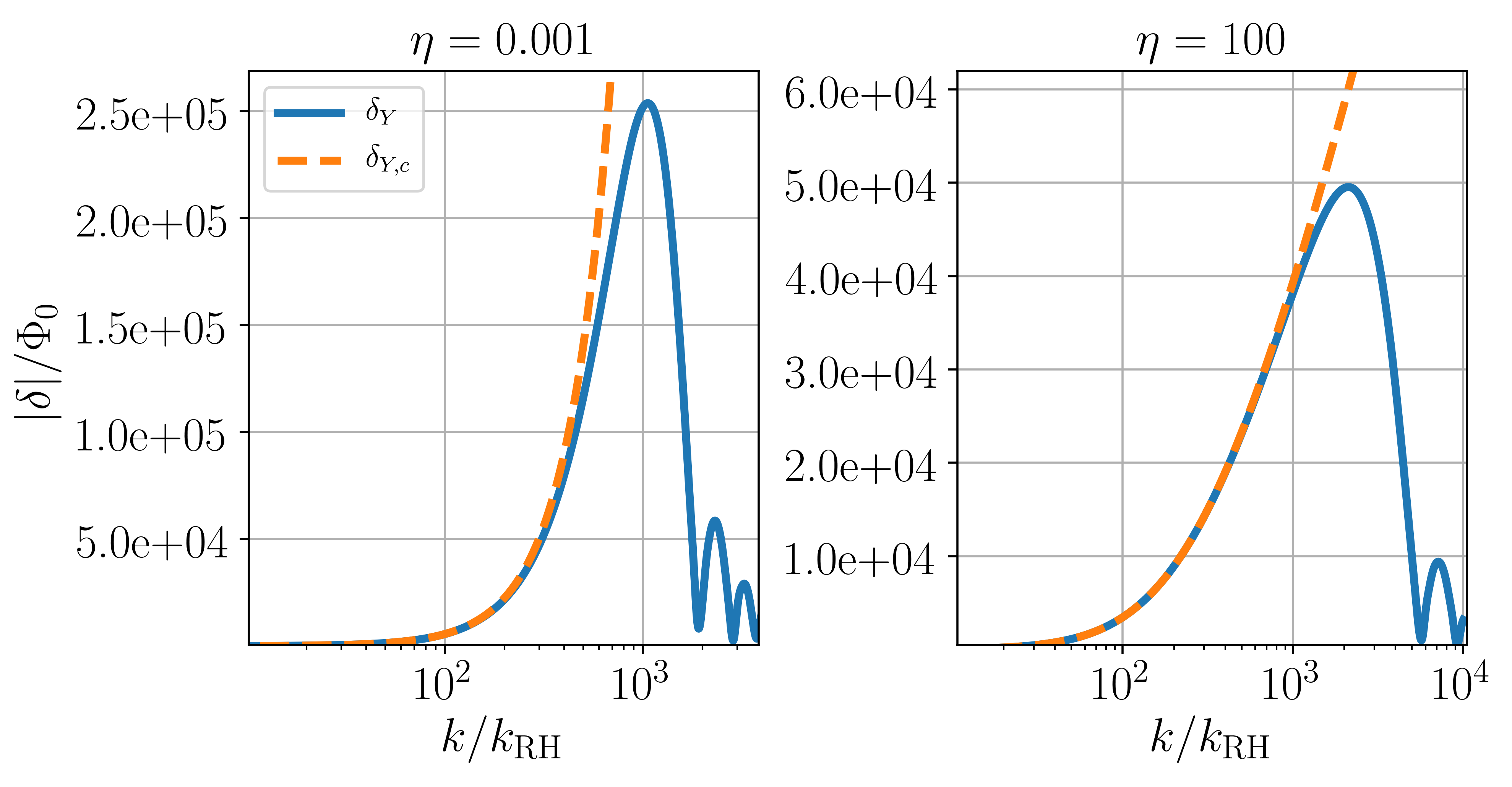}
	\caption{Comparing the amplitude of the first and second peaks in $\delta_Y(k)$. For both $\eta<1$ and $\eta>1$, the second peak amplitude $\lesssim 0.25 \delta_Y(k_{\rm pk})$.}
	\label{fig-deltaylin}
\end{figure}

We wish to fit a functional form to $T(k)$ that accurately models the transfer function. As can be seen in Figure~\ref{fig-deltaylin}, the oscillatory pattern in $\delta_Y(k)$ for $k>k_{\rm pk}$ has a much lower amplitude than $\delta_Y(k_{\rm pk})$: $\delta_Y$ at the second peak scale is $\lesssim 0.25 \delta_Y(k_{\rm pk})$ for both $\eta<1$ and $\eta>1$. Since perturbations at the scale of the first peak will collapse long before modes on smaller scales, we do not expect perturbations with $k>k_{\rm pk}$ to significantly affect the microhalo population. We will therefore prioritize accurately describing the first peak in $\delta_Y(k)$, while neglecting the smaller peaks at $k>k_{\rm pk}$. The function that best fits $T(k)$ and accurately describes the first peak in $\delta_Y(k)$ is \begin{equation} \label{tkfit}
	T(k) = \exp \left[-\left(\frac{k}{k_{\mathrm{cut}}}\right)^n\right] \,,
\end{equation} where both $n$ and $k_{\rm cut}$ are fitting parameters.

\subsection{Scenarios With Initially Subdominant $Y$ particles}
\label{tf-rd}

\begin{figure}[!htb]
	\centering
	\includegraphics[scale=1]{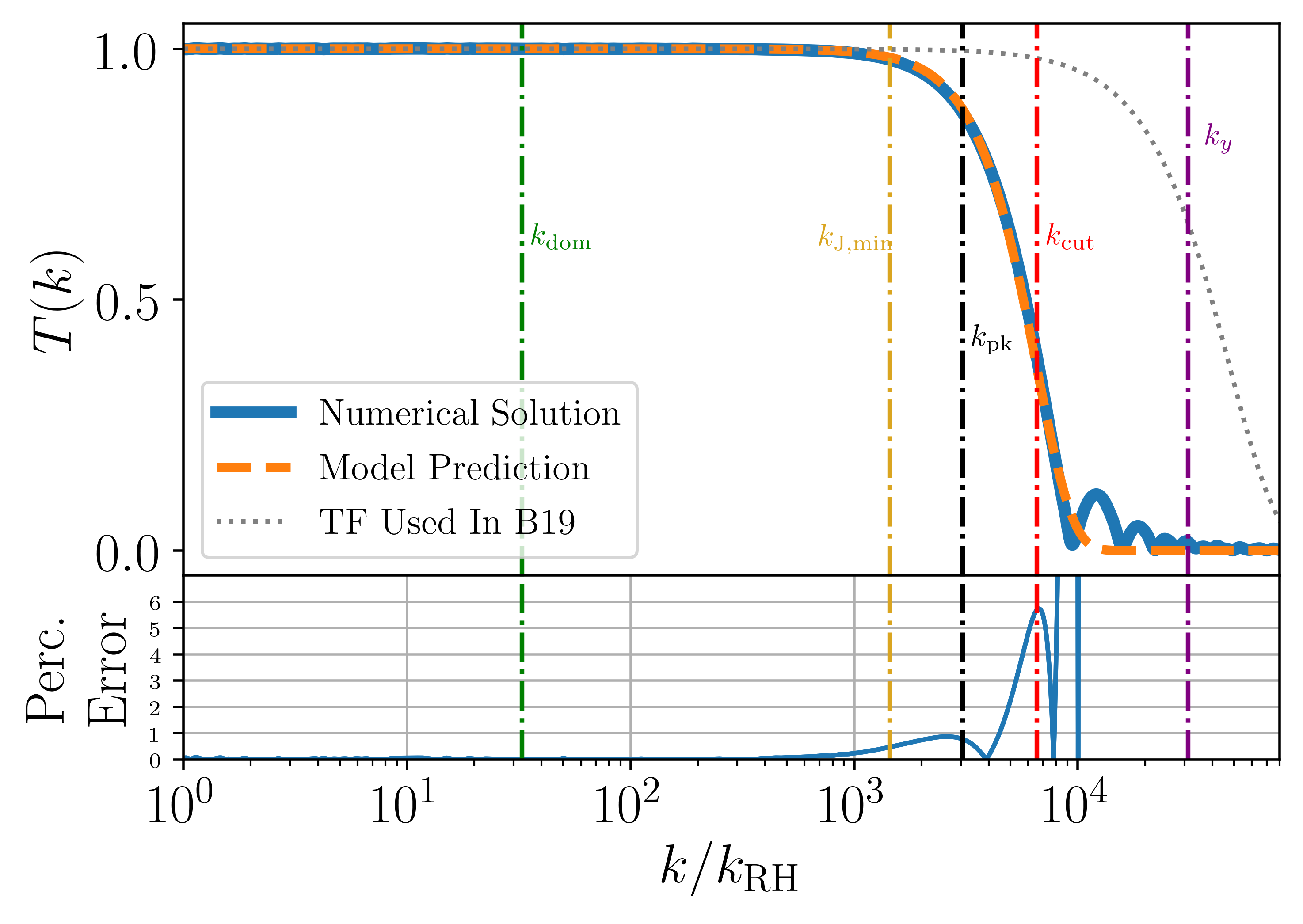}
	\caption{Transfer function $T(k) \equiv \delta_Y(k) / \delta_{Y,c}(k)$, evaluated at $\rh{a}$ for $m=2$ TeV, $\rh{T} = 20$ MeV and $\eta = 500$. The dashed orange curve shows $T(k)$ given by our model, with $k_{\rm cut}$ given by Eq.~(\ref{kc-rd}) and $n = 2.7$.  The transfer function $\exp [-k^2/(2k_y^2)]$ used in B19 is shown by the grey dotted curve, where $k_y$ is the wavenumber corresponding to the horizon when $m = T_{\rm hs}$, given by Eq.~(\ref{kykrh}). The bottom panel shows the percentage error between $T(k)$ and our model prediction. }
	\label{fig-tofk}
\end{figure}

Figure~\ref{fig-tofk} shows $T(k)$ for $m = 2$ TeV, $\rh{T} = 20$ MeV and $\eta = 500$. The transfer function equals unity for $k \lesssim k_{\rm J,min}$. As $k$ increases beyond $k_{\rm J,min}$, $T(k)$ falls off in amplitude as $\delta_Y$ is increasingly suppressed relative to $\delta_{Y,c}$. After the fall-off, $T(k)$ shows oscillations in $k$ that reflect the small-scale decaying oscillations of $\delta_Y(k)$ due to the pressure of the $Y$ particles. Figure~\ref{fig-tofk} also shows the transfer function used in B19, given by $\exp [-k^2 / (2k_y^2)]$, and we see that our transfer function falls off at comparatively smaller $k$ values.

The dashed orange curve in Figure~\ref{fig-tofk} shows the fit to $T(k)$ using the function given by Eq.~(\ref{tkfit}), with fit parameters $k_{\rm cut} / \rh{k} = 6539$ and $n = 2.7$. The bottom panel shows the percentage error between $T(k)$ and the fitting function. At $k=k_{\rm pk}$, the value of the fitting function is within 1\% of the numerical value of $T(k)$. 

\begin{figure}[!htb]
	\centering
	\includegraphics[]{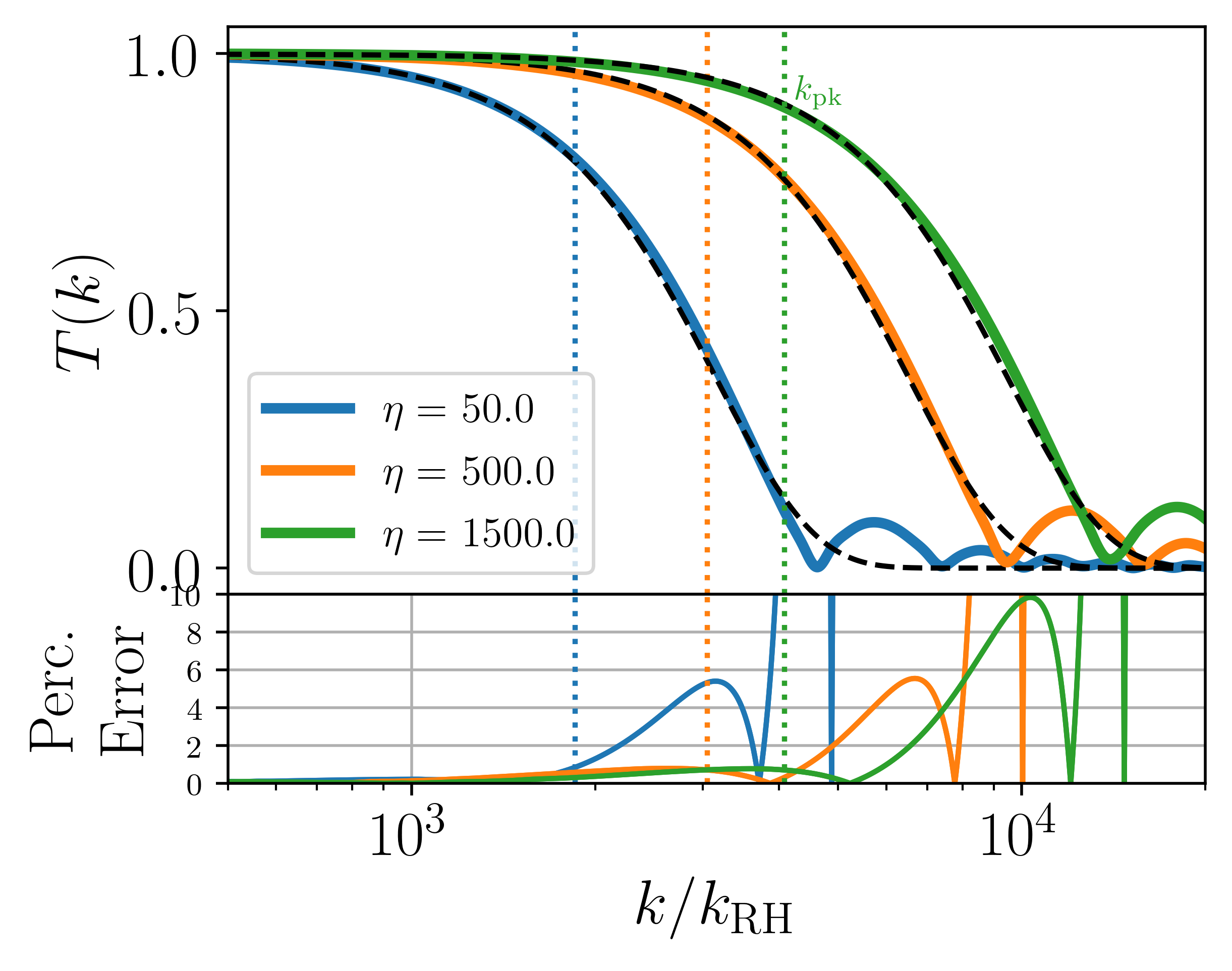}
	\caption{\textbf{Top}: The functional form $T(k) = \exp [-(k/k_{\mathrm{cut}})^{2.7}]$ (dashed black lines) compared to the numerically evaluated $T(k) = \delta_Y(k) / \delta_{Y,c}(k)$ at $\rh{a}$ for three different values of $\eta$. The dotted vertical lines mark $k_{\rm pk}$. The value of $k_{\rm cut}$ is given by Eq.~(\ref{kc-rd}). \textbf{Bottom}: Percentage error between the numerical $T(k)$ and the best fit functions. }
	\label{fig-tk3}
\end{figure}

For $\eta$ between 3 and 1500, the best-fit values for $n$ are between 2.60 and 2.78. Since this variation is small for a range of $\eta$ that spans nearly two orders of magnitude, we fix $n=2.7$ for $\eta > 1$. 
With $n$ fixed, we derive an expression for the cut-off scale $k_{\rm cut}$ by relating it to the peak scale evaluated in Sec.~\ref{pk-rd}.
For $k \geq 10 k_{\rm dom}$, the numerical solutions for $\delta_{Y,c}(k)$ follow a logarithmic function of $a_{\rm dom}/a_k$, where the mode $k$ enters the horizon at $a_k$:
\begin{equation} \label{dyc-log}
	\delta_{Y,c}(k > 10 k_{\rm dom},\rh{a}) = 4.86 \Phi_0 \left(\frac{k_{\rm dom}}{\rh{k}} \right)^2 \ln \left( 0.21 \left[ \frac{g_{*}(T(a_k))}{g_{*}(T_{\rm dom})} \right]^{\frac{1}{6}} \frac{k}{k_{\rm dom}} \right),
\end{equation} where Eq.~(\ref{kkd}) was used to express $a_{\rm dom}/a_k$ in terms of $k/k_{\rm dom}$. Differentiating $\delta_Y(k) = \delta_{Y,c}(k) T(k)$ with respect to $k$, while ignoring the weak $k$-dependence of $g_{*}(T(a_k))$ and using the expression for $T(k)$ given by Eq.~(\ref{tkfit}) with $n=2.7$ generates a relation between $k_{\rm cut}$ and the peak wavenumber $k_{\rm pk}$ that maximizes $\delta_Y(k)$: \begin{equation}\label{kcut}
	\frac{k_{\mathrm{cut}}}{k_{\mathrm{pk}}} = \left[ 2.7 \log \left( 0.21 \left[ \frac{g_{*}(T(a_{\rm pk}))}{g_{*}(T_{\rm dom})} \right]^{\frac{1}{6}}\frac{k_{\mathrm{pk}}}{k_{\mathrm{dom}}} \right) \right] ^{\frac{1}{2.7}}.
\end{equation} Using Eqs.~(\ref{kpk-kd}) and (\ref{rdef}) and the expression for $k_{\rm pk}/k_y$ from Eq.~(\ref{kpk-rd}), we have \begin{equation}\label{kc-rd}
	\frac{k_{\rm cut}}{k_y} = \frac{\alpha}{b} \sqrt{\frac{\eta}{1 + \eta}} \left[ \frac{g_{*}(T(a_y)) g_{*}(T_{\rm dom})}{g_{*}^2 (T(a_{\rm pk}))} \right]^\frac{1}{6}  W^{-\frac{3}{2}} (0.77 \bar{g}^{\frac{1}{9}} \eta^{\frac{2}{3}}) [ \ln \{ 0.18 \bar{g}^{\frac{1}{6}} \eta W^{-\frac{3}{2}} (0.77 \bar{g}^{\frac{1}{9}} \eta^{\frac{2}{3}}) \} ]^{0.37},
\end{equation} where $\bar{g} = g_{*}^2(T_i) / g_{*}(T_{\rm dom}) g_{*}(T(a_{\rm pk}))$, $W$ is the Lambert W-function and $b$ is 2.70 if the $Y$ particles are bosons and 3.15 if they are fermions. The coefficient $\alpha$ accounts for the slight difference in the peak scale values for $Y$ particles following different statistics, as described in section~\ref{pk-rd}: $\alpha = 1.82$ and 1.84 for bosonic and fermionic $Y$ particles, respectively.  

The numerical solutions for three values of $\eta$ are shown in Figure~\ref{fig-tk3} along with the curves given by Eq.~(\ref{tkfit}) with $n=2.7$ and $k_{\rm cut}$ calculated using Eq.~(\ref{kc-rd}).  The percentage errors between the functional forms and $T(k)$ are plotted in the bottom panel. For $\eta = 1500$, which was the maximum value of $\eta$ for which $T(k)$ was computed, the functional form with $n=2.7$ is within 1\% of the numerical value of $T(k)$ at $k_{\rm pk}$. The percentage error remains less than 8\% for $T(k) \geq 0.25$.  The numerically determined $k_{\rm cut}$ is within 4\% of the analytical expression given by Eq.~(\ref{kc-rd}) for $\eta>50$. 

For smaller values of $\eta$, $k_{\rm cut} \lesssim 10 k_{\rm dom}$, and the expression given by Eq.~(\ref{dyc-log}) becomes increasingly inaccurate. In addition, our analytical prediction of $k_{\rm pk}$ diverges from the peak wavenumber in the range $\eta<50$. Thus, the prediction of $k_{\rm cut}$ given by Eq.~(\ref{kc-rd}) becomes inaccurate. For $1.1 \leq \eta \leq 50$, we empirically find that a power law describes the variation of $k_{\rm cut}/k_y$ with $\eta$:\begin{equation}\label{kc-rd-f}
	\frac{k_{\rm cut}}{k_y} = \frac{\alpha + 0.15}{b} \eta ^{-0.21}.
\end{equation} This expression predicts the cut-off scale to within 2.5\% error for $1.1 \leq \eta \leq 50$. 

\begin{figure}
	\centering
	\includegraphics[width = \textwidth]{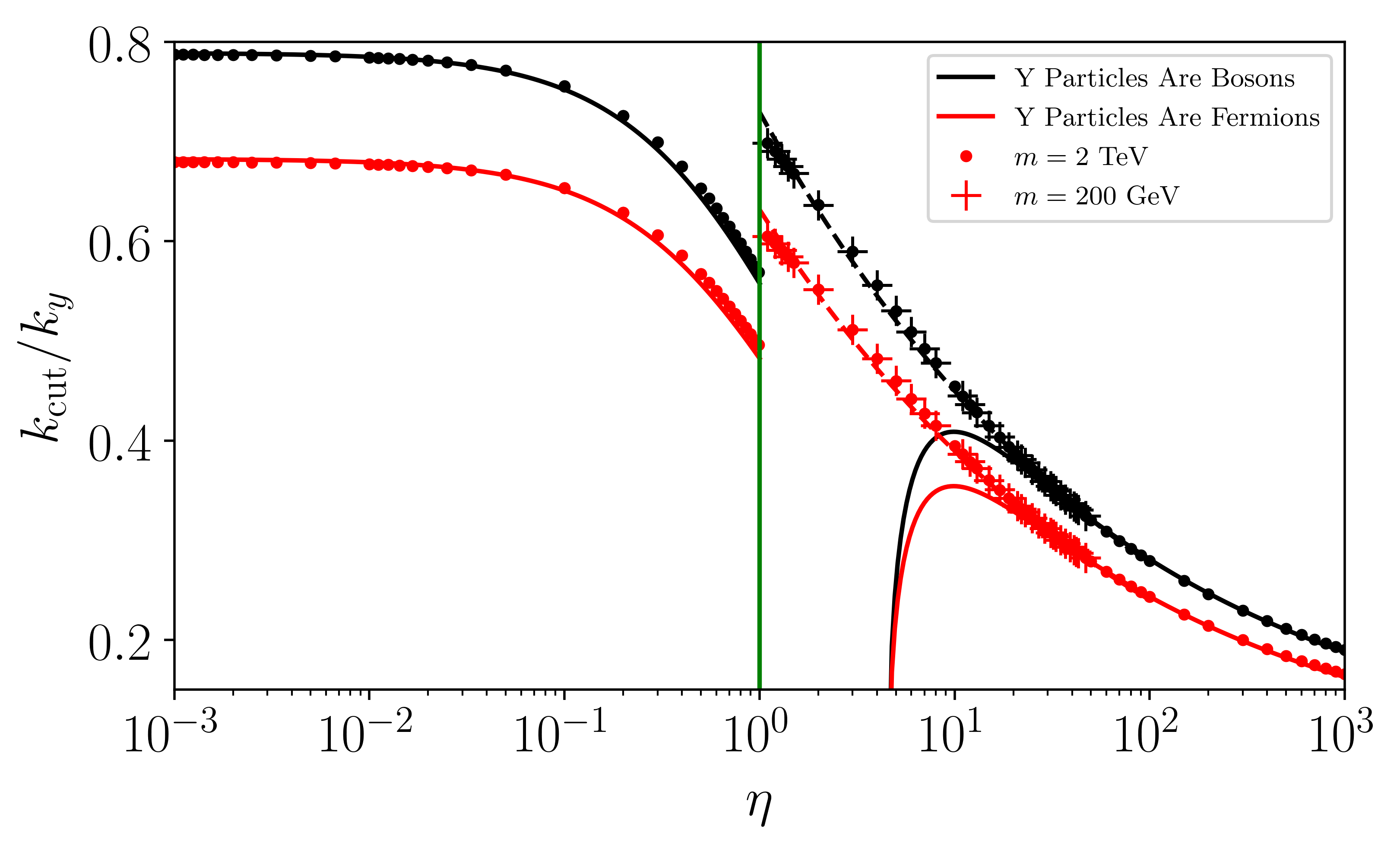}
	\caption{The cut-off scale wavenumber for the transfer function, plotted as $k_{\rm cut}/k_y$ against $\eta$, the initial value of $\rho_R/\rho_Y$. The different colors indicate Bose-Einstein or Fermi-Dirac statistics for the $Y$ particles. The solid curves for $\eta>50$ show the analytical predictions for $k_{\rm cut}/k_y$ given by Eq.~(\ref{kc-rd}), whereas the dashed curves show the power law fits given by Eq.~(\ref{kc-rd-f}) for $1<\eta<50$. The solid curves for $\eta<1$ show the analytical predictions given by Eq.~(\ref{kc-yd}). The dots and crosses represent the $k_{\rm cut}$ determined by fitting the functional form $\exp[-(k/k_{\rm cut})^n]$, with crosses showing cases with $m=200$ GeV and dots showing $m=2$ TeV for the range $1<\eta<50$. This plot assumes $\gs{T} = g_{*S}(T) = 100$.}
	\label{fig-kcut}
\end{figure}

In the $\eta>1$ regime in Figure~\ref{fig-kcut} we show the numerically determined $k_{\rm cut}$ divided by $k_y$ as points plotted for different values of $\eta$ for cases when the $Y$ particles are bosons (red) and fermions (black). For $\eta>50$, the expressions given by Eq.~(\ref{kc-rd}) are plotted as the solid curves. For $1 < \eta < 50$, the power law fits given by Eq.~(\ref{kc-rd-f}) are plotted as the dashed curves. The numerically determined $k_{\rm cut} / k_y$ values are shown for $m = 2$ TeV (dots) and $m = 200$ GeV (crosses). The overlap of the dots and crosses demonstrates that the validity of the power law fit of Eq.~(\ref{kc-rd-f}) is independent of $m$.

\subsection{Scenarios With Initial $Y$-Domination}
\label{tf-yd}

If $\eta<1$, the universe is initially dominated by the energy density of the $Y$ particles, and $\delta_{Y,c}(k,\rh{a}) \propto \rh{a} /a_k \propto  (k/\rh{k})^2$ for all $k \gtrsim \rh{k}$. From our numerical solutions, we find \begin{equation} \label{dyc-yd}
	\delta_{Y,c}(k,\rh{a}) = 0.62 \Phi_{0m} \left( \frac{k}{\rh{k}}\right)^2.
\end{equation} Here, $\delta_{Y,c}$ is evaluated by setting the initial conditions outlined in Appendix \ref{pertsol}, and $\Phi_{0m} = 9\Phi_0/10$ is the primordial metric perturbation in a matter-dominated universe.  This expression for $\delta_{Y,c}(k,\rh{a})$ is used to evaluate $T(k)$ when $\eta<1$. 

We fit the functional form $\exp [-(k/k_{\rm cut})^n]$ to the transfer function, treating $n$ and $k_{\rm cut}$ as free parameters. From fitting $T(k)$ for $10^{-3} < \eta < 1$, the $\eta$-dependence of $n$ can be summarized as \begin{equation} \label{npred-yd}
	n = \begin{cases}
		2.2 - 0.29(\eta - 0.1) & 0.1 \leq \eta < 1\\
		2.2 & \eta < 0.1.
	\end{cases}
\end{equation} The value of $n$ falls with increasing $\eta$ in the range $0.1 \leq \eta < 1$ as the contribution of the SM radiation density to the Hubble rate becomes increasingly significant. To find an analytical expression for $k_{\rm cut}$, we use Eq.~(\ref{dyc-yd}) with $T(k) = \exp [-(k/k_{\rm cut})^n]$ and maximize $\delta_Y(k) = T(k) \delta_{Y,c}(k)$ with respect to $k$ to obtain the peak wavenumber $k_{\rm pk}$ in terms of $k_{\rm cut}$: \begin{equation} \frac{k_{\rm cut}}{k_{\rm pk}} = \left(\frac{n}{2}\right)^{1/n} . \end{equation} Substituting $k_{\rm pk}/k_y$ from Eq.~(\ref{kpk-yd}), we have \begin{equation}\label{kcky-yd}
	\frac{k_{\rm cut}}{k_y} = \frac{\gamma}{b} \left(\frac{n}{2}\right) ^{\frac{1}{n}} \frac{1}{\sqrt{1 + \eta}},
\end{equation} where $\gamma = 2.055$ and 2.065 for bosonic and fermionic $Y$ particles, respectively, and once again, $b =2.70$ if the $Y$ particles are bosons and $b=3.15$ if they are fermions.

For $\eta<0.1$, $n = 2.2$, which can be substituted in Eq.~(\ref{kcky-yd}) to obtain $k_{\rm cut} / k_y =1.04\gamma/(b \sqrt{1 + \eta})$. As $\eta$ increases from 0.1 to 1, $n$ decreases linearly and $k_{\rm pk}/k_y$ increases relative to the prediction of Eq.~(\ref{kpk-yd}). Since the cut-off scale follows the relation $k_{\rm cut} \propto k_{\rm pk} (n/2)^{1/n}$, the rise of $k_{\rm pk}/k_y$ nearly cancels out the effect of $n$ decreasing. As a result, 
\begin{equation} \label{kc-yd}
	\frac{k_{\rm cut}}{k_y} = \frac{1.04\gamma}{b} \frac{1}{\sqrt{1 + \eta}}.
\end{equation}
predicts $k_{\rm cut}$ to within 2\% error even for $0.1<\eta<1$. 

Figure~\ref{fig-kcut} shows $k_{\rm cut}/k_y$ in the range $\eta<1$ for the cases when the $Y$ particles are bosons (red) and fermions (black). The solid lines for $\eta<1$ show the predictions of Eq.~(\ref{kc-yd}). The discontinuity at $\eta=1$ between the analytical predictions given by Eqs.~(\ref{kc-rd-f}) and (\ref{kc-yd}) arises because Eq.~(\ref{dyc-yd}) for $\delta_{Y,c}(k)$ was used to evaluate $T(k)$ for cases with $\eta<1$. Unlike the $\delta_{Y,c}$ used in $T(k)$ for $\eta>1$, Eq.~(\ref{dyc-yd}) neglects the contribution of the SM radiation density to the Hubble rate.

\begin{figure}[h!]
	\centering
	\includegraphics[width = \textwidth]{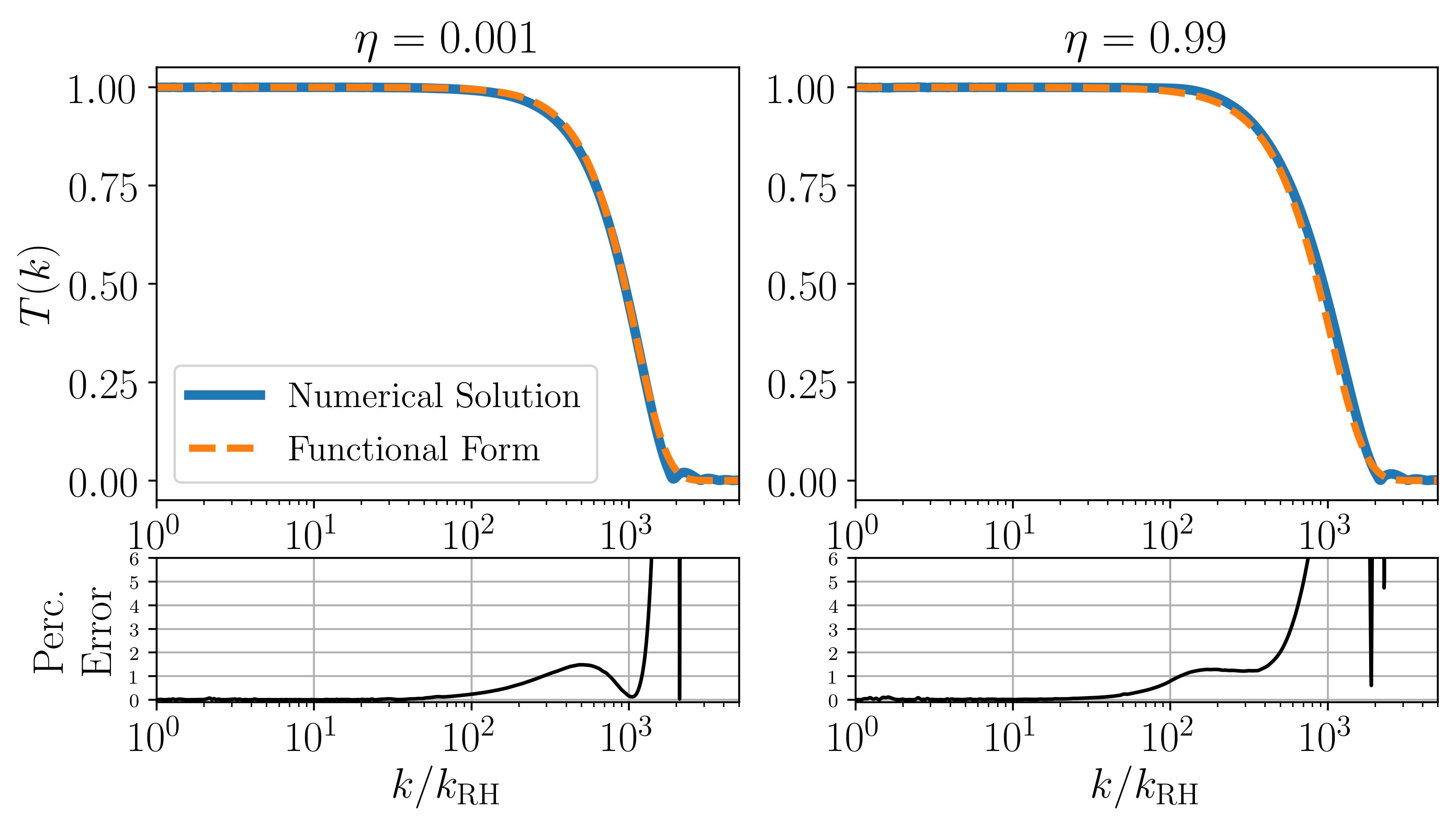}
	\caption{\textbf{Top}: Numerically obtained transfer functions $T(k)$ (solid blue lines) and curves given by $\exp [-(k/k_{\rm cut})^n]$ (dashed orange lines) for the cases $\eta=0.001$ (left) and $\eta=0.99$ (right). For the dashed curves, $k_{\rm cut}$ is given by Eq.~(\ref{kc-yd}) and $n$ by Eq.~(\ref{npred-yd}).  \textbf{Bottom}: Percentage error between the numerical $T(k)$ and the curves given by our models. }
	\label{fig-tkfit-yd}
\end{figure}

Figure \ref{fig-tkfit-yd} shows $T(k)$ and the functional form $\exp[-(k/k_{\rm cut})^n]$ with $k_{\rm cut}$ given by Eq.~(\ref{kc-yd}) and $n$ by Eq.~(\ref{npred-yd}), for the cases $\eta=0.001$ (left panel) and $\eta=0.99$ (right panel). The bottom panels show the percentage error between $T(k)$ and the functional forms. 

\section{The Peak Amplitude and Observational Prospects}
\label{peakampl}

Having determined the transfer functions and the cut-off scale, we can estimate how the EMDE impacts the dark matter annihilation rate today.  Following the same procedure as Ref.~\cite{blanco19}, we use the Press-Schechter formalism \cite{press-schechter} to obtain the abundance of microhalos, and then we calculate the annihilation rate per volume assuming that the microhalos have an NFW profile with concentration $c=2$ at their formation time. The increase of the annihilation rate due to microhalo formation is quantified by the boost factor $B(z) \equiv \langle \rho_X^2 \rangle / \langle \rho_X \rangle ^2 - 1$.  The resulting boost factor initially increases with time as more halos form, but then it starts to decrease as the earliest-forming microhalos are absorbed into larger halos and their Press-Schechter abundance decreases.  The first microhalos are very dense and are expected to survive their absorption into larger halos \cite{Delos:2019lik,Delos:2019tsl,Shen:2022ltx}, so we take the maximum value of $B(z)$ to be the boost factor today ($B_0$).  

\begin{figure}[!htb]
\centering
\includegraphics[scale=0.9]{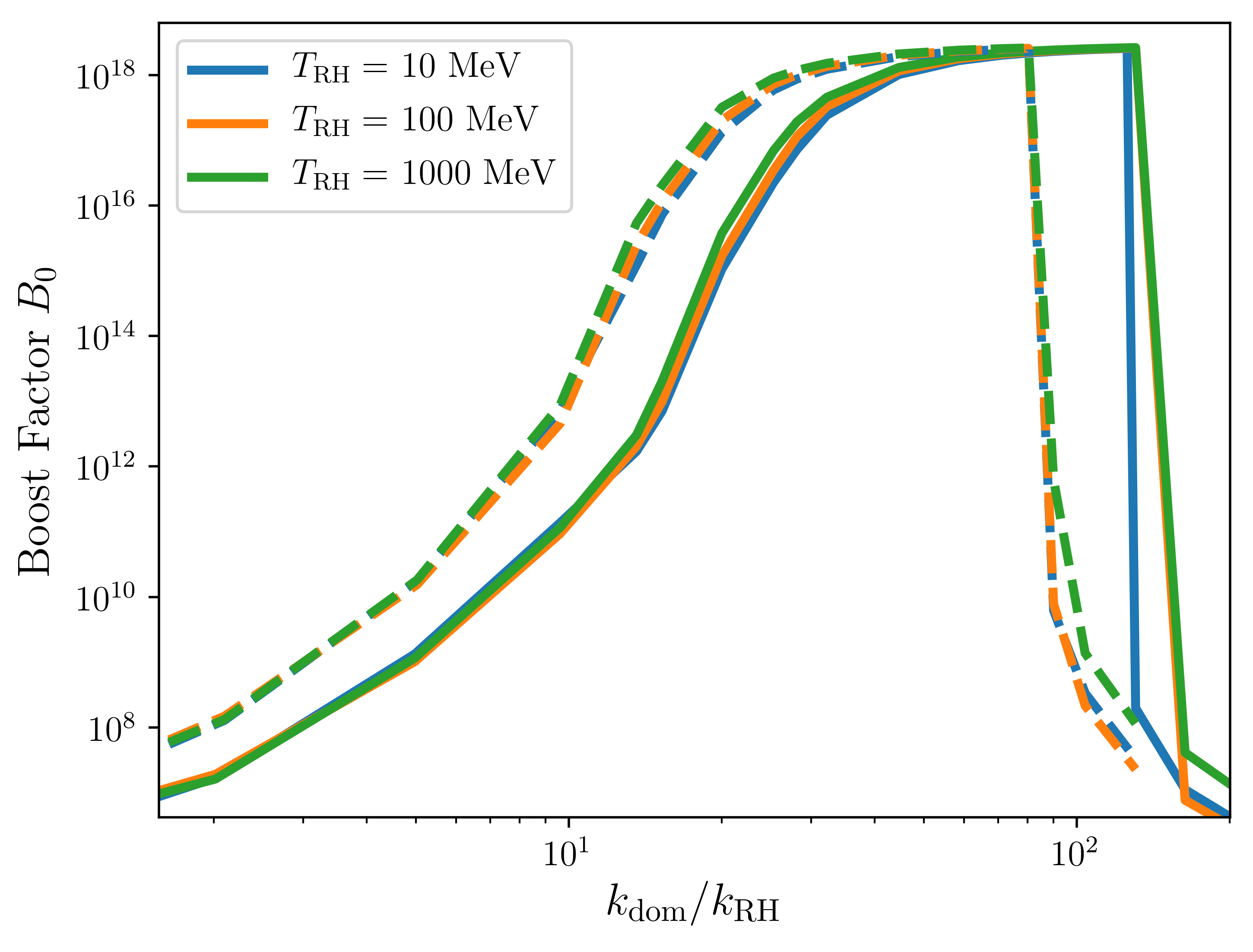}
\caption{The boost to the DM annihilation signal as a function of the duration of the EMDE, quantified by $k_{\rm dom}/\rh{k}$. The solid lines show the boost if the transfer functions given by this work are used, while the dashed lines show the boost if the transfer functions from Ref.~\cite{blanco19} are used. Using our transfer functions increases the value of $k_{\rm dom}/\rh{k}$ that produces a given value of $B_0$ because our transfer functions suppress perturbations on larger scales, so a longer EMDE is required to generate the same enhancement to the matter power spectrum.}
\label{fig-b0}
\end{figure}

Figure~\ref{fig-b0} shows $B_0$ as a function of $k_{\rm dom}/\rh{k}$ for $\eta = 12.69$, which corresponds to a hidden sector containing vector $Y$ particles ($g=3$) and Dirac fermion $X$ particles ($g_X = 4$) that kinetically decoupled from the SM particles while the $Y$ and $X$ particles were relativistic. 
The solid lines show $B_0$ calculated using the transfer functions given by this work for three $\rh{T}$ values, while the dashed lines show $B_0$ for the same reheat temperatures calculated using the $\exp[-k^2/(2k_y^2)]$ transfer function assumed in Ref.~\cite{blanco19}. As the duration of the EMDE increases, earlier structure formation leads to higher boost factors.  The growth of density perturbations during the EMDE can even lead to halo formation prior to matter-radiation equality.  Following Ref.~\cite{blanco19}, we only consider halos that form at redshifts less than $10^6$, and this restriction is responsible for the plateau at $B_0 \simeq 2\times10^{18}$.
The boost increases with increasing $\rh{T}$ at fixed $k_{\rm dom}/\rh{k}$ due to a longer period of logarithmic growth for $\delta_X$ after the EMDE.  
 
If the EMDE is long enough, microhalos form before reheating.  These microhalos are dominated by $Y$ particles, so they dissipate when the $Y$ particles decay~\cite{blanco19}. The released dark matter particles have randomly oriented velocities with magnitudes boosted by nonlinear structure formation. This gravitational heating imposes a free-streaming cut-off on the power spectrum after reheating that reduces the boost factor to the standard $\Lambda$CDM prediction, $B_0 \sim 10^6$. We implement gravitational heating following the ``optimistic'' approach from Ref. \cite{blanco19}: a free-streaming cut-off is applied to the matter power spectrum based on the minimum virial velocity of the halos that contain 20\% of the dark matter at the end of the EMDE, and no free-streaming cut-off is imposed if less than 20\% of the dark matter is bound into halos during the EMDE. The sharp decrease in $B_0$ due to gravitational heating can be seen at $k_{\rm dom} / \rh{k}\simeq 100$ in Figure \ref{fig-b0}. The slight red tilt of the primordial power spectrum causes the reduction of the boost due to gravitational heating to move to higher values of $k_{\rm dom}/\rh{k}$ for higher $\rh{T}$.

Figure~\ref{fig-b0} shows that $B_0$ for $k_{\rm dom} / \rh{k} \lesssim 80$ is smaller for our transfer functions compared to those from Ref.~\cite{blanco19}; our transfer functions suppress longer-wavelength perturbations, which reduces the amplitude of the peak in the power spectrum and delays the formation of bound structures. For \mbox{$k_{\rm dom} / \rh{k} \gtrsim 80$}, the comparatively fewer microhalos predicted by our transfer functions at reheating implies that the reduction of the boost due to gravitational heating happens for larger $k_{\rm dom}/\rh{k}$ (longer EMDEs) compared to when the transfer functions from Ref.~\cite{blanco19} are used. 

Figure \ref{fig-b0} demonstrates that the annihilation boost does not strongly depend on $\rh{T}$, but it is highly sensitive the duration of the EMDE, which sets the peak amplitude of the matter power spectrum for a fixed value of $\eta$. While the peak scale controls the size of the first microhalos that form during or after an EMDE, the peak amplitude determines their formation times because gravitational collapse occurs when $\delta \simeq 1.68$.  The central density of a halo forming at $a_f$ scales as $a_f^{-3}$. Consequently, structures form earlier and have denser cores if the power spectrum has a higher peak \cite{obs_blinov}, which yields larger annihilation boosts up to the point that the peak becomes high enough that halos form during the EMDE.

\begin{figure}
	\centering
	\includegraphics[width = 0.7\textwidth]{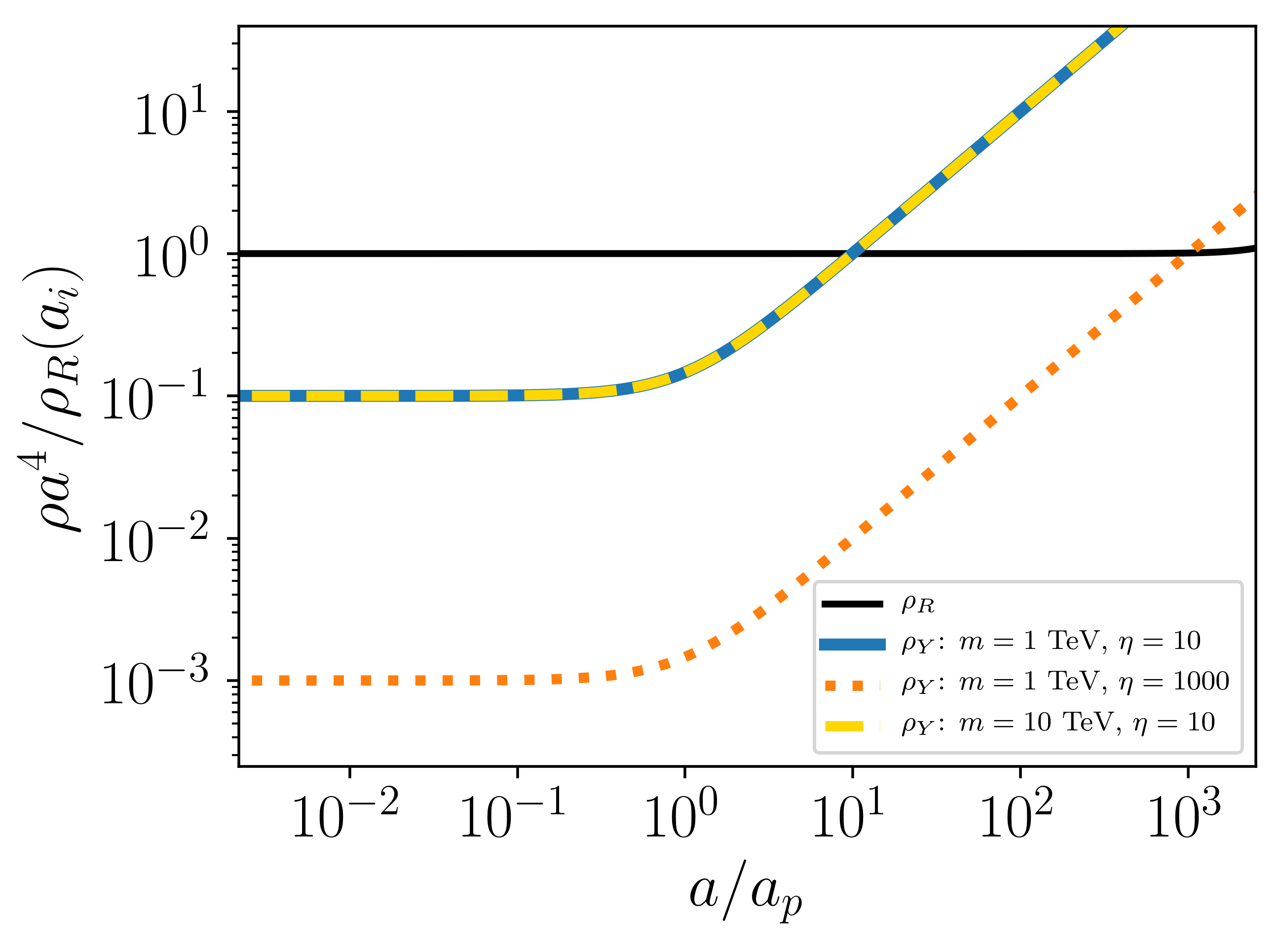}
	\caption{The effect of varying $m$ and $\eta$ on the evolution of SM and $Y$ densities. The solid blue, dashed yellow and dotted orange lines show $\rho_Y a^4$ as a function of $a/a_p$, where $a_p$ is the scale factor at which the $Y$ particles can be considered to have become nonrelativistic. The EMDE begins when the respective lines cross the horizontal black line, which shows $\rho_R a^4$. The blue and orange lines have the same $m = 1$ TeV but different $\eta$; $a_{\rm dom} / a_p$ differs for these cases. In contrast, the solid blue and dashed yellow lines have the same $a_{\rm dom} / a_p$ even with different values of $m$ since $\eta$ is the same.  }
	\label{fig-backdiff}
\end{figure}

If the $Y$ particles initially dominate the universe, the amplitude of the peak in the matter power spectrum depends only on the duration of EMDE: a longer EMDE implies a longer period of linear perturbation growth, translating to a higher peak in the power spectrum.  If the $Y$ particles are initially subdominant, then the peak amplitude also depends on how long the universe remains radiation dominated after the $Y$ particles become nonrelativistic.  Figure \ref{fig-backdiff} demonstrates that $a_{\rm dom} / a_p$ depends exclusively on $\eta$: $a_{\rm dom} / a_p$ remains the same if $m$ is varied while $\eta$ is held fixed.  
We use the transfer functions derived in the previous section to calculate $\delta_{\rm pk} \equiv \delta_Y(k_{\rm pk})$ and evaluate observational prospects in terms of $\eta$ and the duration of the EMDE.

If the $Y$ particle is initially subdominant ($\eta>1$), the following fitting function describes $\delta_{Y,c}(k,\rh{a})$ well for $k > 10\rh{k}$: \begin{equation} \label{dyc-rd}
    \delta_{Y,c} (k > 10\rh{k},\rh{a}) =  0.596\Phi_0  \left(\frac{k_{\rm dom}}{k_{\rm RH}}\right)^2  \frac{ \ln (1 + 0.22q)}{0.22q} q^2 p(q) ,
\end{equation} where $q = k/k_{\rm dom}$ and $p(q) = [1 + 1.11q  + (0.94q)^2 +  (0.63q)^3 + (0.45q)^4]^{-1/4}$. The peak amplitude is $\delta_Y(k_{\rm pk}) = \delta_{Y,c}(k_{\rm pk}) T(k_{\rm pk})$, where $T(k) = \exp [-(k/k_{\rm cut})^{2.7}]$. Equation~(\ref{dyc-rd}) shows that $\delta_{Y,c}(k_{\rm pk})$ is separable into $(k_{\rm dom}/k_{\rm RH})^2$, which sets the duration of the EMDE, times a function of $k_{\rm pk} / k_{\rm dom}$. The ratio $k_{\rm pk} / k_{\rm dom}$ depends only on $\eta$ and the $Y$ particle statistics, as shown by Eq.~(\ref{kpk-kd}). Furthermore, $k_{\rm pk}/k_{\rm cut}$ also depends only on $\eta$ and the statistics of the $Y$ particles, as illustrated by Eqs.~(\ref{kc-rd}) and (\ref{kc-rd-f}).   
The peak scale is given by Eq.~(\ref{kpk-rd}), and Eq.~(\ref{kc-rd}) provides $k_{\rm cut}$ for $ \eta > 50$. For $1 < \eta < 50$, the power-law fit of Eq.~(\ref{kc-rd-f}) gives $k_{\rm cut}$, and the peak scale is well-described by the fit
\begin{equation}
    \frac{k_{\rm pk}}{k_y} = \frac{\alpha - 0.18}{b} \eta^{-0.299},
\end{equation} where $\alpha=1.82$ and $b = 2.70$ for bosonic $Y$ particles and $\alpha =1.84$ and $b=3.15$ for fermionic $Y$ particles.  This prescription for calculating $\delta_{\rm pk}$ matches the numerically determined maximum of $\delta_Y(k)$ to within 4\%.  As expected, $\delta_{\rm pk} \equiv \delta_Y(k_{\rm pk})$ depends on $k_{\rm dom}/k_{\rm RH}$ and $\eta$.

\begin{figure}[htb]
	\centering
	\includegraphics[width = \textwidth]{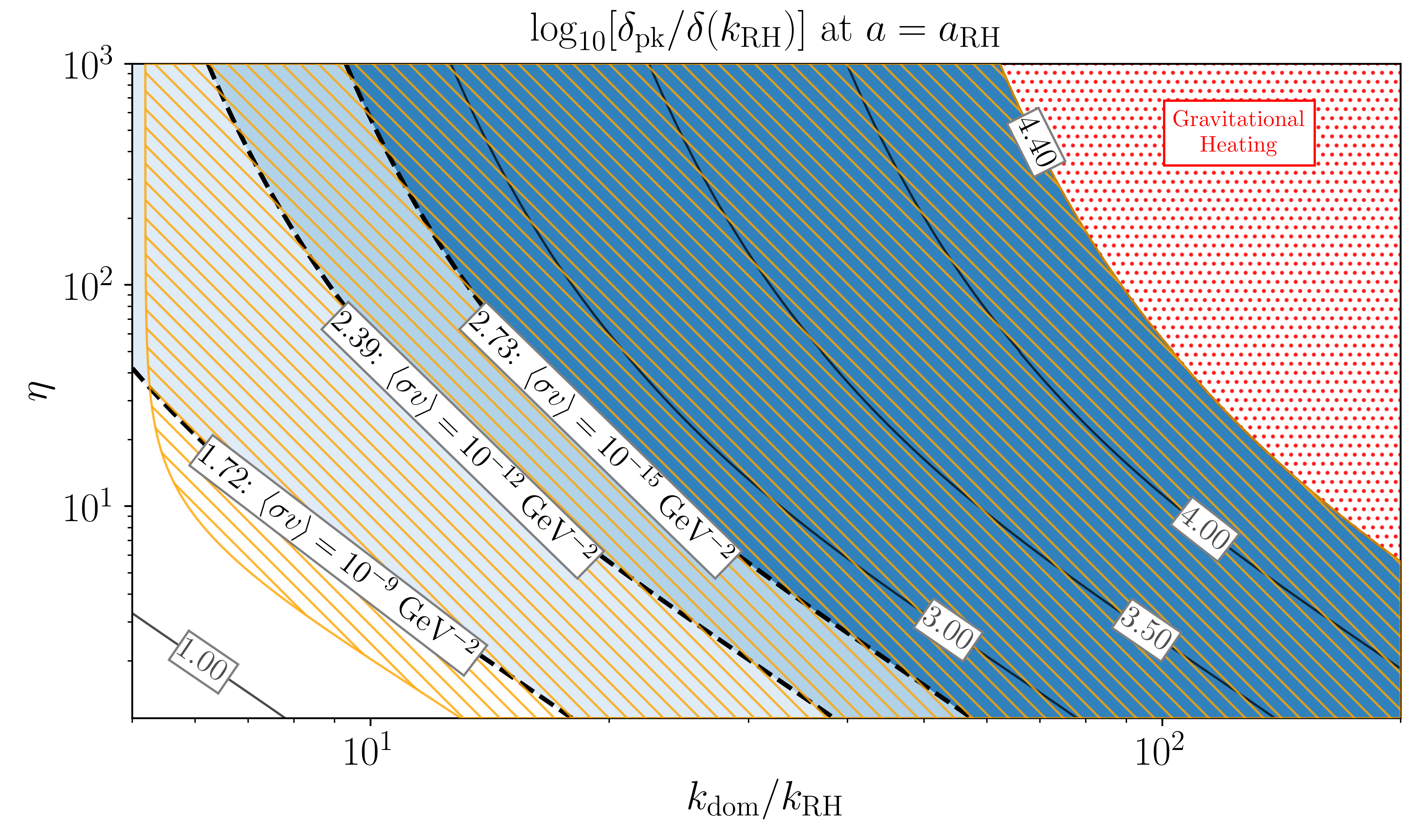}
	\caption{Contours of $\log_{10} [ \delta_{\rm pk} / \delta(\rh{k})]$, the maximum enhancement to $\delta_Y$ at the end of the EMDE, as a function of $\eta$ and $k_{\rm dom} / \rh{k}$ for cases in which the universe is initially dominated by SM radiation. This plot takes $g_{*}(T) = 100$. The red dotted region shows cases in which microhalos are erased at the end of the EMDE and gravitational heating suppresses structure formation after the EMDE. The thick dashed contours mark the largest $k_{\rm dom} / \rh{k}$ values that are compatible with IGRB observations for three values of the DM annihilation cross-section and $m_X = 10^{6}$ GeV, with the labels showing the $\langle \sigma v \rangle$ value considered. For instance, the white region to the left of the dashed contour with $\log_{10} [ \delta_{\rm pk} / \delta(\rh{k})] = 1.72$ is allowed for $\langle \sigma v \rangle \geq 10^{-9} \text{ GeV}^{-2}$. The deep blue region is excluded if $\langle \sigma v \rangle \geq 10^{-15} \text{ GeV}^{-2}$.  The yellow hatched region can be probed with pulsar timing arrays with 100 pulsars observed weekly for 25 years for cases with $\rh{T} \lesssim 20$ MeV.}
	\label{fig-peakheight-rd}
\end{figure}

Figure~\ref{fig-peakheight-rd} shows $\delta_{\rm pk} / \delta_Y(\rh{k})$ at $\rh{a}$ for a bosonic $Y$ particle, where $\delta_Y(\rh{k},\rh{a}) \simeq 3.05 \Phi_0$ is the value of $\delta_Y$ at the scale at which the power spectrum begins deviating from the power spectrum in scenarios without an EMDE, and we have continued to neglect the scale-dependence of $\Phi_0$. Since $\delta_Y(k)\simeq \delta_X(k)$, and $\delta_X(k)$ only logarithmically increases with $k$ for modes that enter the horizon during radiation domination, this ratio nearly equals the maximum enhancement to the power spectrum. Figure~\ref{fig-peakheight-rd} shows that $\delta_{\rm pk}$ increases with increasing $k_{\rm dom} / \rh{k}$, as expected.  Figure~\ref{fig-peakheight-rd} also shows that $\delta_{\rm pk}$ is rather sensitive to $\eta$ for $\eta \lesssim 10$, but that sensitivity wanes as $\eta$ increases. As can be seen in Figure~\ref{fig-deltayk}, $\delta_Y(k)$ continues to rise steeply with $k$ for $k \gtrsim k_\mathrm{dom}$ and plateaus when $k \gtrsim 10k_\mathrm{dom}$.  
Therefore, $\delta_{\rm pk}$ sharply depends on $k_\mathrm{pk}$ if $k_\mathrm{pk} \lesssim 10 k_\mathrm{dom}$ but then becomes less sensitive to $k_\mathrm{pk}$ as increasing $\eta$ increases $k_\mathrm{pk}/k_\mathrm{dom}$.  The variation of the peak height with $\eta$ thus becomes weaker, as indicated by the contours in Figure~\ref{fig-peakheight-rd} becoming increasingly vertical as $\eta$ increases.

If the $Y$ particles dominate the universe before they become nonrelativistic ($\eta<1$), the expression for $\delta_{Y,c}$ from Eq.~(\ref{dyc-yd}) implies that  
\begin{equation} \label{dpk-yd}
    \delta_{\rm pk} = 0.62 \Phi_{0m} \left( \frac{k_y}{\rh{k}}\right)^2 \left( \frac{k_{\rm pk}}{k_y}\right)^2 \exp \left[-\left(\frac{k_{\rm pk}}{k_{\rm cut}}\right)^n \right],
\end{equation} where $k_{\rm pk}/\rh{k}$ was split into $(k_y/\rh{k}) (k_{\rm pk}/k_y)$. The expressions for $k_{\rm pk} / k_y$ and $k_{\rm cut} / k_y$ can be taken from Eqs.~(\ref{kpk-yd}) and (\ref{kc-yd}) respectively, while $n$ is given by Eq.~(\ref{npred-yd}). The peak amplitude can be predicted to within 4\% error for $\eta < 1$ using these expressions. 

\begin{figure}[htb]
	\centering
	\includegraphics[width = \textwidth]{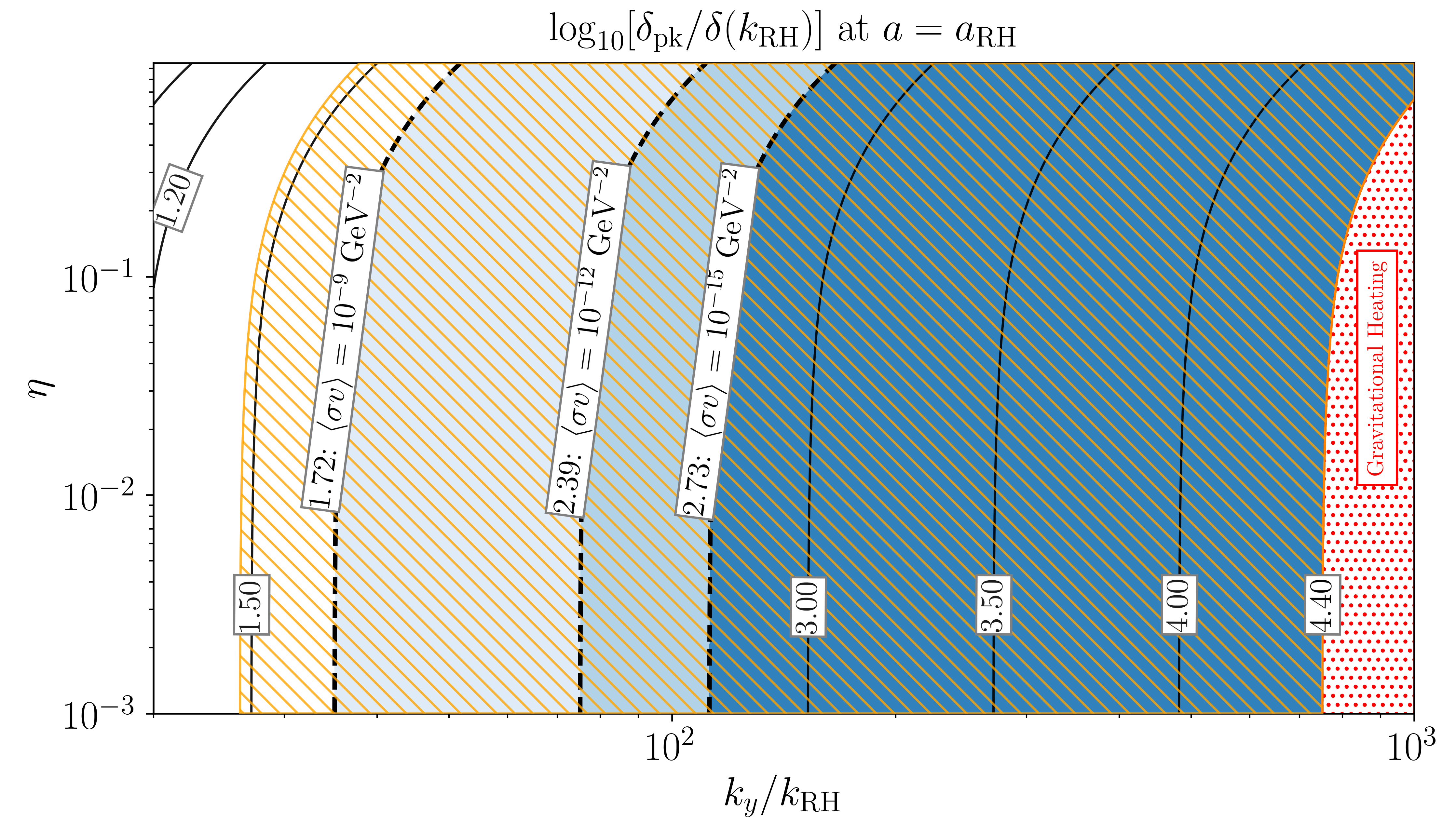}
	\caption{Contours of $\log_{10} [ \delta_{\rm pk} / \delta(\rh{k})]$, the maximum enhancement to $\delta_Y$ at the end of the EMDE, as a function of $\eta$ and $k_y / \rh{k}$ for cases in which the universe is initially dominated by the $Y$ particles. The red dotted region shows cases in which microhalos form during the EMDE and gravitational heating suppresses structure formation after the EMDE. The thick dashed contours show the limits of the parameter space compatible with IGRB observations for three values of the DM annihilation cross-section and $m_X = 10^{6}$ GeV, with the labels showing the $\langle \sigma v \rangle$ value considered. The yellow hatched region can be probed with pulsar timing arrays with 100 pulsars observed weekly for 25 years for cases with $\rh{T} \lesssim 20$ MeV.}
	\label{fig-peakheight-yd}
\end{figure}

Equation~(\ref{dpk-yd}) shows that $\delta_{\rm pk}$ is proportional to $(k_y / \rh{k})^2$; this ratio sets the duration of the EMDE. Apart from $k_y/\rh{k}$, the other factors in Eq.~(\ref{dpk-yd}) depend only on $\eta$ and the $Y$ particle statistics. For $\eta < 0.1$, the $\eta$-dependence becomes negligible as the universe becomes increasingly $Y$-dominated at the time of horizon entry of the peak scale. Figure \ref{fig-peakheight-yd} shows the peak enhancement to $\delta_Y$ at the end of the EMDE as a function of $k_y / \rh{k}$ and $\eta$ for a bosonic $Y$ particle. The contours show that $\delta_{\rm pk}$ increases with $k_y/\rh{k}$ as the EMDE becomes longer and that the enhancement is independent of $\eta$ for $\eta<0.1$.

For very long EMDEs, $B_0$ falls below $10^8$ due to gravitational heating (as can be seen in Figure \ref{fig-b0}). All scenarios with higher power spectrum peaks will be similarly affected by the destruction of bound structures during reheating. In Figures \ref{fig-peakheight-rd} and \ref{fig-peakheight-yd}, the red dotted regions indicate points with $\delta_{\rm pk} / \delta(\rh{k}) \gtrsim 10^{4.4}$, which is the peak enhancement for $\{ \eta, k_{\rm dom} / \rh{k} \} = \{12.69,150\}$. These areas mark the parameter combinations for which gravitational heating reduces the abundance of microhalos after the EMDE.

If dark matter is a thermal relic, then it is possible to constrain EMDE cosmologies using limits on the dark matter annihilation rate. Reference \cite{blanco19} calculated an annihilation rate per mass in a given volume if the dark matter resides within a population of microhalos: \begin{equation} \label{gammab0}
    \frac{\Gamma_{\rm DM}}{M_X} = \frac{1}{2} \left(\frac{\langle \sigma v \rangle/m_X^2}{\text{GeV}^{-4}} \right)B_0 \times (8.098 \times 10^{-47}) \times \Omega_X h^2,
\end{equation} where $\langle \sigma v \rangle$ is the velocity-averaged DM annihilation cross-section and $M_X$ is the total dark matter mass in the volume. Since the annihilation rate in these scenarios is proportional to the number density of microhalos and thus the dark matter density, the annihilation signal produced is similar to that from decaying dark matter.  Assuming that annihilation and decay events both produce two primary particles, the annihilation rate can be related to an effective DM lifetime by equating the rate of particle production from annihilation, $2M_X (\Gamma_{\rm DM} / M_X)$, to the particle production rate from decaying DM of mass $2m_X$ and lifetime $\tau$, given by $(2/\tau)[M_X / (2m_X)]$. This yields $\tau_{\rm eff}^{-1} = 2m_X (\Gamma_{\rm DM}/M_X)$, which should be compared to the bounds on the lifetime of DM particles with mass $2m_X$. 

Using Fermi-LAT observations of the Isotropic Gamma Ray Background (IGRB) \cite{igrb}, Ref. \cite{taueff} established $\tau_{\rm eff} \gtrsim 10^{28}$ seconds for a variety of decay channels and DM masses ranging from 10 GeV to $10^9$ GeV. By connecting $B_0$ to peak height, we can estimate the allowed regions of the EMDE parameter space based on this lower limit on $\tau_{\rm eff}$. Using $\Omega_X h^2 = 0.12$ and $m_X = 10^6$ GeV, the IGRB constraint on $\tau_{\rm eff}$ translates to $B_0 \lesssim 10^{10}$ for $\langle \sigma v \rangle = 10^{-9} \text{ GeV}^{-2}$, which is close to the canonical WIMP DM cross-section of $2.2 \times 10^{-26} \text{ cm}^3 \text{ s}^{-1}$ \cite{dmsigmav}. From Figure~\ref{fig-b0}, we see that $\{ \eta, k_{\rm dom} / \rh{k} \} = \{12.69,7\}$ corresponds to $B_0 = 10^{10}$. The white unshaded region to the bottom left in Figure \ref{fig-peakheight-rd} marks all points with $\delta_{\rm pk}$ values that are smaller than $\delta_{\rm pk}$ for $\{ \eta, k_{\rm dom} / \rh{k} \} = \{12.69,7\}$, marking the allowed parameter space based on this constraint. The white unshaded region to the left in Figure \ref{fig-peakheight-yd} marks the allowed area of the $\eta$-$k_y/\rh{k}$ space for cases with $\eta<1$ based on the same peak height limits.  

If the DM particle freezes out during or before the EMDE, then a smaller annihilation cross section is required to generate the observed DM density \cite{hid4,allahverdi2019,guidice2000}. Lowering the cross-section increases the upper limit on $B_0$, expanding the allowed parameter space. Considering $\langle \sigma v \rangle = 10^{-12} \text{ GeV}^{-2}$, the light blue regions to the left of the thick dashed contour line with $\log_{10} [ \delta_{\rm pk} / \delta(\rh{k})] = 2.39$ are allowed in addition to the white regions in Figures \ref{fig-peakheight-rd} and \ref{fig-peakheight-yd}. With $\langle \sigma v \rangle = 10^{-15} \text{ GeV}^{-2}$, the allowed region increases to include the medium blue regions to the left of the contour lines with $\log_{10} [ \delta_{\rm pk} / \delta(\rh{k})] = 2.73$, with only the deep blue regions excluded in both plots. 

The annihilation contours in Figures \ref{fig-peakheight-rd} and \ref{fig-peakheight-yd} are presented as estimates because they assume that the relation between $B_0$ and $\delta_{\rm pk}$ for a single value of $\eta$ can be extended to other $\eta$ values.  
Power spectra with the same peak heights have different peak scales for different values of $\eta$, but the fact that $B_0$ is largely insensitive to $T_{\rm RH}$ indicates that $B_0$ does not depend strongly on $k_{\rm pk}$. 
Changing $\eta$ also changes the shape of the power spectrum around its peak.  The impact of peak shape on $B_0$ has not been extensively studied, but the values for $B_0$ computed in Ref.~\cite{sten_gr} for a power spectrum with $\eta \ll 1$ and the $B_0$ values for $\eta=12.69$ in Figure \ref{fig-b0} differ by less than an order of magnitude for power spectra with the same $\delta_{\rm pk}$.  We conclude that the $B_0-\delta_{\rm pk}$ relation for $\eta = 12.69$ provides a strong indication of which $\delta_{\rm pk}$ values can be ruled out by limits on the dark matter annihilation rate.

It is also possible to detect the microhalos that form after an EMDE through their gravitational influence.  Pulsar timing arrays (PTAs) are promising probes of EMDE cosmologies \cite{pta1,pta_sten}: PTAs are sensitive to both the Shapiro time delays as signals pass through microhalos and the Doppler shifts that result when a microhalo pulls on a pulsar, with the latter being most sensitive to sub-earth-mass microhalos \cite{Ramani:2020hdo}. With weekly observations and an RMS timing residual of 10 ns, Ref. \cite{pta_sten} showed that microhalos resulting from EMDE-enhanced power spectra with $\rh{T} \lesssim 20$ MeV and $k_{\rm cut} / \rh{k} > 20$ can be detected at $2\sigma$ significance if 100 pulsars are observed for 25 years or if 1000 pulsars are observed for 15 years. 

Reference \cite{pta_sten} used power spectra from initially $Y$-dominated EMDEs \cite{emde2} with a Gaussian cut-off. A cut-off given by $k_{\rm cut} / \rh{k} = 20$ on their power spectra implies that $\delta(k)$ peaks at around $24 \rh{k}$ with a value close to $ 27\delta(\rh{k})$. Consequently, power spectra with $\delta(24 \rh{k}) \gtrsim 27 \delta(\rh{k})$ will produce microhalos that have similar detection prospects to those produced by initially $Y$-dominated EMDE scenarios with $k_{\rm cut} / \rh{k} \gtrsim 20$. Such cases are marked by the yellow hatched regions in Figures \ref{fig-peakheight-rd} and \ref{fig-peakheight-yd}; we expect that these EMDE scenarios with $\rh{T} \lesssim 20$ MeV will generate signals that are detectable by the PTAs described above. 

Another possible method of observing the microhalos resulting from an EMDE comes from how they impact the magnification of stars that pass behind the lensing caustics of galaxy clusters \cite{caustic1,caustic2,caustic3}.   As the star passes through the caustic, fluctuations in the dark matter density generate variations in the star's brightness, which can be used to detect sub-earth-mass microhalos.
Reference \cite{obs_blinov} identified the ranges of microhalo masses and central densities that can be detected using this method by imposing lower bounds on the magnitude of the observed brightness fluctuations and on the abundance of microhalos.  They demonstrated that microhalos that meet their detection criteria are generated by
 a power spectrum that rises as $k^4$ for $\rh{k} < k < k_{\rm pk}$ and decreases sharply for $k > k_{\rm pk}$, reaching a peak enhancement of $10^4$ times the $\Lambda$CDM power spectrum at $k_{\rm pk} \approx 10^{8.5} k_{\rm eq}$, where $k_{\rm eq}$ is the horizon scale at matter-radiation equality. 
Employing our transfer functions, the parameters $\rh{T} = 4$ MeV with $\eta = 0.01$ and $k_y / \rh{k} = 100$ generate a power spectrum with a $k^4$ rise before a peak at the scale $k_{\rm pk} \approx 10^{8.5} k_{\rm eq}$. The peak enhancement is $\delta_{\rm pk} / \delta(\rh{k}) = 126$, corresponding to a power spectrum peak enhancement factor of $\approx 10^4$. Using $\rh{T} = 4$ MeV, $\eta = 8$ and $k_{\rm dom} / \rh{k} = 20$ generates a power spectrum with a similar peak scale and peak enhancement. Although the scaling is not strictly $k^4$ for $\rh{k} < k < k_{\rm pk}$ in this case, the power spectrum is logarithmic for only a narrow range of $k$ near the peak, making this power spectrum roughly similar to one that rises as $k^4$ before the peak. For similar peak enhancements, power spectra with a $k^4$ rise before the peak and larger $k_{\rm pk}$ values from cases with $\rh{T}$ up to a few hundred MeV also result in microhalos that can be detected using caustic microlensing observations \cite{obs_blinov}.

\section{Summary and Discussion}
\label{summ}

The linear growth of dark matter perturbations during an early matter-dominated era (EMDE) leads to the formation of microhalos much earlier than in standard cosmologies \cite{emde2,emde1,ae15}.  These dense microhalos may be detected gravitationally by upcoming pulsar timing arrays \cite{pta1,obs_blinov,pta_sten} and through their impact on stellar microlensing events in galaxy clusters \cite{caustic1,caustic2,caustic3,obs_blinov}.  They can also boost the dark matter annihilation rate by several orders of magnitude \cite{ae15,blanco19,sten_gr}.  Perturbation growth is suppressed for modes that enter the horizon while the particle that dominates the universe during the EMDE has significant relativistic pressure.  The DM power spectrum manifests this suppression as a small-scale cut-off, which strongly affects the DM annihilation signal \cite{ae15,sten_gr}. The small-scale cut-off also impacts the prospects of detecting the structures formed in EMDE cosmologies via pulsar timing arrays \cite{pta1,obs_blinov,pta_sten} and caustic microlensing \cite{obs_blinov}. It is therefore important to accurately calculate this cut-off scale, so that EMDE scenarios with initially hot hidden sectors may be tested against observational data. In this paper, we have investigated the small-scale cut-off in the matter power spectrum that results from the relativistic initial state of the particle responsible for the EMDE.

We employed a custom Boltzmann solver to calculate the evolution of perturbations in a universe with an initially relativistic hidden sector particle ($Y$). We found that the evolution of subhorizon perturbations in the $Y$ particle density ($\delta_Y$) depends on the wavelength of the perturbation mode compared to a time-varying Jeans length. This Jeans length is set by the sound speed of the $Y$ particles, and it increases while they are relativistic and then starts decreasing after they transition to nonrelativistic behavior. As long as the Jeans length is greater than a perturbation mode's wavelength, $\delta_Y$ oscillates, while it grows when the Jeans length drops below the mode wavelength. Therefore, linear growth during the EMDE starts later for smaller-scale modes. This suppression of growth due to relativistic pressure generates a peak in the power spectrum of $\delta_Y$: for wavelengths smaller than the peak scale, the power spectrum falls off in amplitude due to the delayed onset of growth during the EMDE, whereas longer wavelength modes have less time to grow during the EMDE because they enter the horizon later. This peak is inherited by the dark matter power spectrum as dark matter particles fall into the gravitational wells created by the clustered $Y$ particles during the EMDE. 

To describe how the relativistic pressure of the $Y$ particles affects the matter power spectrum, we provided transfer functions that relate the matter perturbations of initially cold and hot hidden sectors. These transfer functions generate the matter power spectrum following an EMDE arising from an initially hot hidden sector without the cumbersome calculation of the density evolution of the hidden sector particle as it transitions from relativistic to nonrelativistic behavior. The transfer functions take the form $\exp [-(k/k_{\rm cut})^n]$, where $n$ depends on $\rho_{\rm SM}/\rho_Y$ when the $Y$ particles were relativistic ($\eta$) and $k_{\rm cut}$ is the cut-off scale. We found that $k_{\rm cut}/k_y$ is a function of $\eta$ and the $Y$ particle statistics, where $k_y$ is the wavenumber of the mode that enters the horizon when the hidden sector temperature equals the $Y$ particle mass $m$. The ratio $k_y / \rh{k}$, where $\rh{k}$ is defined as the horizon wavenumber at the end of the EMDE, depends on $\eta$ and is proportional to $(m / \rh{T})^{2/3}$. We found that $k_{\rm cut}$ is smaller than $k_y$, which was used as an estimate of the cut-off scale in Ref.~\cite{blanco19}. Our result also disproves the claim in Ref.~\cite{zhang} that the horizon scale at the start of the EMDE sets the cut-off scale. 

The cut-off scale determines the power spectrum peak height, which sets the formation times and central densities of the first microhalos. The peak height $\delta_{\rm pk}$ depends on the EMDE duration and $\eta$. Longer EMDEs translate to larger $\delta_{\rm pk}$ since they involve longer periods of linear perturbation growth. For $\eta<1$, $\delta_{\rm pk} \propto (k_y / \rh{k})^2$. For $\eta>1$, $\delta_{\rm pk} \propto (k_{\rm dom} / \rh{k})^2$, where $k_{\rm dom}$ is the horizon wavenumber at the start of the EMDE. If $\eta<0.1$, the peak height is independent of $\eta$ because the subdominant SM radiation density does not affect the evolution of perturbations prior to the end of the EMDE. For $\eta>1$, the peak height depends on $\eta$ because $\eta$ determines how long it takes the $Y$ particle to dominate the universe after it becomes nonrelativistic.  Relating the peak height to $\eta$ and the EMDE duration enables the discussion of observational prospects and constraints in the parameter space of hidden-sector EMDE histories. 

If the peak is high enough for microhalos to form during the EMDE, the evaporation of these microhalos at reheating causes the ejection of DM particles at high speeds in random directions. This gravitational heating leads to a free-streaming cut-off on the power spectrum after the EMDE. The exact evolution of this free-streaming cut-off and its relation to the abundance of microhalos that formed during the EMDE is unknown, with recent studies \cite{remnants} even suggesting that the remnants of evaporated halos may re-collapse into bound structures around the epoch of matter-radiation equality. We identified the regions of parameter space where 20\% or more of the dark matter is gravitationally heated; the affected parameter space has peak enhancement $\delta_{\rm pk} / \delta(\rh{k}) \gtrsim 10^{4.4}$. This corresponds roughly to cases with $\eta^{1/4} k_{\rm dom} / \rh{k} \gtrsim 250$ for $1 < \eta \lesssim 1000$ and $k_y / \rh{k} \gtrsim 800$ for $\eta<1$.  

Since the microhalos that form after an EMDE track the dark matter density, the annihilation rate within microhalos can be compared to the rate of particle production from decaying dark matter to define an effective DM lifetime. We used constraints on the dark matter lifetime \cite{taueff} based on the Fermi-LAT observations of the Isotropic Gamma Ray Background (IGRB) \cite{igrb} to derive bounds on the dark matter annihilation boost $B_0$.  By connecting the bounds on $B_0$ to the peak height, we identified the allowed regions of the parameter space of hidden-sector EMDE histories. Assuming a DM mass of $10^6$ GeV with an annihilation cross-section close to the canonical value of $10^{-9} \text{ GeV}^{-2}$, the IGRB constraint allows cases obeying $\eta^{1/3}k_{\rm dom} / \rh{k} \lesssim 25$ for $\eta>1$, or cases with $k_y / \rh{k} \lesssim 35$ for $\eta<1$. Smaller cross-sections are required to match the currently observed DM relic abundance if the DM freezes out during or before an EMDE; the allowed parameter space expands for these lower cross-section values and for higher values of DM mass. Since $k_{\rm cut} < k_y$, our transfer functions yield less structure formation for the same EMDE duration compared to Ref. \cite{blanco19}. For cases not involving gravitational heating, we therefore obtain smaller annihilation boost factors for the same EMDE duration. In addition, this reduced structure formation also delays the onset of gravitational heating, which happens for longer EMDEs compared to Ref. \cite{blanco19}.  

We also found that a large portion of the parameter space of hidden-sector EMDEs can be probed with the pulsar timing arrays discussed in Ref.~\cite{pta_sten}. For example, weekly observations of 100 pulsars for 25 years would detect microhalos generated from EMDEs with $\rh{T} \lesssim 20$ MeV,  $ 30\lesssim k_y / \rh{k} \lesssim 800$, and $\eta < 0.1$, where the upper limit on $k_y/ \rh{k}$ comes from the uncertainty associated with the disruption to the post-EMDE power spectrum due to gravitational heating.  If $\eta>1$, the same PTA observations would detect microhalos resulting from EMDEs with $13 \lesssim \eta^{1/4}k_{\rm dom} / \rh{k} \lesssim 250$ and $\rh{T} \lesssim 20$ MeV.  Furthermore, EMDE power spectra for reheat temperatures less than ${\cal O}$(100 MeV) with peaks that are enhanced by a factor of $10^4$ relative to the standard $\Lambda$CDM power spectrum lead to microhalos that produce detectable brightness fluctuations when stars pass through the lensing caustics of galaxy clusters \cite{obs_blinov}.  

Our calculation of the small-scale power spectrum cut-off that results from the relativistic pressure yields a more accurate mapping between the properties of EMDE cosmologies and the observable signals that can help detect or constrain them.  Our work thus improves our ability to probe the microscopic properties of hidden sectors and the expansion history of the early universe. 

\acknowledgments

We thank M.~Sten Delos for helpful discussions and Alexander Sobotka, A.~Turchaninova (AT), and Hwan Bae for useful feedback on the paper draft.  K.J.M. and H.G. are supported by NSF Grant AST-2108931. A.L.E. is supported in part by NSF CAREER grant PHY-1752752.

\appendix

\section{The Evolution Of The Homogeneous Hidden Sector Background}
\label{ysol-method}

The $Y$ particles that dominate the energy density of the universe during the EMDE are initially relativistic and transition to a pressureless state as the hidden sector temperature decreases. This appendix presents calculations for the evolution of the equation of state, pressure, and density of the $Y$ particles.

\subsection{Method}

We use energy conservation and number density conservation to formulate a system of coupled differential equations for quantities related to the hidden sector temperature $T_{\rm hs}$ and the chemical potential of the $Y$ particles, denoted by $\mu$. We will assume here that the $Y$ particle has $g$ degrees of freedom and write the energy density $\rho_Y$, pressure $P_Y$, and number density $n_Y$ as thermodynamic integrals:

\begin{subequations}
\begin{align}
\rho_Y(T_{\rm hs},\mu) &= \frac{g}{2 \pi^2} \int^{\infty}_{0} \frac{E(p) }{e^{(E- \mu)/T_{\rm hs}}  \pm 1} p^2 dp \,; \label{rhoy-th} \\
P_Y (T_{\rm hs},\mu) &= \frac{g}{6 \pi^2} \int^{\infty}_{0} \frac{p^2}{E(p)}\frac{1}{e^{(E- \mu)/T_{\rm hs}} \pm 1} p^2 d p \,; \label{py-th} \\
n_Y(T_{\rm hs},\mu) &= \frac{g}{2 \pi^2} \int^{\infty}_{0} \frac{1}{e^{(E- \mu)/T_{\rm hs}} \pm 1} p^2 dp\,,
\end{align}
\end{subequations} where the $\pm 1$ in the denominator denotes fermions (upper sign) or bosons (lower sign). These integrals can be expressed in terms of dimensionless quantities: $ z \equiv p/T_{\rm hs}$, $x \equiv m/T_{\rm hs}$, $ \Delta \equiv - \mu/T_{\rm hs}$, $\epsilon \equiv \frac{E}{T_{\rm hs}} = \frac{\sqrt{p^2 + m^2}}{T_{\rm hs}} = \sqrt{z^2 + x^2} $ so that \begin{subequations}
\begin{align} 
\rho_Y(x,\Delta) =& \frac{gm^4}{2 \pi^2 x^4} \int^{\infty}_{0} \frac{z^2 \epsilon}{\dis} dz \,; \\
P_Y(x,\Delta) =& \frac{gm^4}{6 \pi^2 x^4} \int^{\infty}_{0} \frac{z^4 \epsilon^{-1}}{\dis} dz \,; \\
n_Y(x,\Delta) =& \frac{gm^3}{2 \pi^2 x^3} \int^{\infty}_{0} \frac{z^2 }{\dis} dz .
\end{align}
\end{subequations} We also introduce the notation \begin{equation}
J[f] \equiv \int^{\infty}_{0} \frac{z^2 f (z,x,\Delta)}{\dis} dz.
\end{equation} 

Conservation of energy density and number density imply\begin{subequations} \label{ec}
\begin{align} 
\dot{\rho}_Y + 3H(1 + w_Y)\rho_Y &= 0 \,, \label{rhoy-eq} \\ 
\dot{n}_Y + 3Hn_Y &= 0 \,,
\end{align}
\end{subequations} where $w_Y(x,\Delta) \equiv P_Y(x,\Delta) / \rho_Y(x,\Delta) $ and overdots denote proper time derivatives. Note that we have ignored the decay of the $Y$ particles because an EMDE only occurs when the $Y$ particles transition to nonrelativistic behavior well before they decay. To transform Eqs.~(\ref{ec}) into differential equations for $x$ and $\Delta$, we express $\dot{\rho}_Y$ and $\dot{n}_Y$ in terms of $\dot{x}$ and $\dot{\Delta}$. For $\rho_Y$, we have \begin{equation}
\dot{\rho}_Y = \frac{gm^4}{2 \pi^2 x^4} \left(R_1 \dot{x}-J\left[\frac{\epsilon e^{(\epsilon + \Delta)}}{\dis}\right] \dot{\Delta} \right),
\end{equation} where
\begin{equation}
R_1 =  -\frac{4}{x} J[\epsilon] + xJ[\epsilon^{-1}] - xJ\left[\frac{e^{(\epsilon + \Delta)}}{\dis}\right].
\end{equation}
And similarly,
\begin{equation}
\dot{n}_Y = \frac{gm^3}{2\pi^2 x^3} \left(N_1 \dot{x} - J\left[\frac{e^{(\epsilon + \Delta)}}{\dis}\right]\dot{\Delta} \right),
\end{equation} where
\begin{equation}
 N_1 = - \frac{3}{x} J[1] - x J \left[\frac{e^{(\epsilon + \Delta)}}{\epsilon(\dis)}\right].
\end{equation} Substituting these definitions in Eqs.~(\ref{ec}) and isolating $\dot{x}$ and $\dot{\Delta}$ yields \begin{subequations}\label{eqs-xd}
\begin{align} 
\dot{x} &= \frac{3HN_0 R_2 - 3H(1+w_Y)R_0 N_2 }{R_1 N_2 - N_1 R_2} \,, \\
\dot{\Delta} &= \frac{3H(1 + w_Y)R_0 N_1 - 3HN_0 R_1}{R_1 N_2 - N_1 R_2}.
\end{align}
\end{subequations}

Equations~(\ref{eqs-xd}) are solved to obtain the hidden sector temperature $T_{\rm hs}$ and the chemical potential $\mu$ of the $Y$ particles as a function of time. With $T_{\rm hs}$ and $\mu$ obtained, the time evolution of the $Y$ particle density and pressure can be calculated using Eqs.~(\ref{rhoy-th}) and (\ref{py-th}) respectively. Finally, the equation of state $w_Y$ and the sound speed $c_{sY}^{2}$ can be computed; $w_Y = P_Y / \rho_Y$ and $c_{sY}^2 = \delta P_Y/ \delta \rho_Y =  P_Y ' / \rho_Y ' = w_Y - w_Y'/(3(1 + w_Y))$, where primes denote $d/d \ln a$.

\subsection{Modeling The Transition From Relativistic to Nonrelativistic Behavior}

The evolution of $\rho_Y$ can be modeled by a broken power law with a pivot point $a_p$.  Since the $Y$ particles become nonrelativistic long before their comoving density is altered by their decays, an expression for $a_p$ can be obtained by conserving $n_Y a^3$ through the transition from relativistic to nonrelativistic behavior.

We use the ansatz $a_p/a_i = bT_{\rm hs,i}/m$, where $T_{\rm hs,i}$ is the hidden sector temperature at $a_i$. Let us assume that the $Y$ particles have become fully nonrelativistic at scale factor $a_{\rm nr}$. Conserving particle number implies $n_Y(a_i) a_i^3 = n_Y(a_{\rm nr}) a_{\rm nr}^3$. Since the $Y$ particles are nonrelativistic at $a_{\rm nr}$, we can write \begin{equation} \label{rdev1}
    \rho_Y(a_{\rm nr}) = m n_Y(a_{\rm nr}) = m n_Y(a_i) \frac{a_i^3}{a_{\rm nr}^3}. 
\end{equation} Using the broken-power-law model for $\rho_Y(a)$, we can also express \begin{equation} \label{rdev2}
    \rho_Y(a_{\rm nr}) = \rho_Y(a_i) \left( \frac{a_i}{a_p} \right)^4 \left(\frac{a_p}{a_{\rm nr}} \right)^3,
\end{equation} where we have used $ \rho_Y(a) \propto a^{-4}$ for $a_i \leq a \leq a_p$ and $\rho_Y(a) \propto a^{-3}$ for $a_p \leq a \leq a_{\rm nr}$. At $a_i$, the $Y$ particles are relativistic with a temperature $T_{\rm hs,i}$, therefore $n_Y(a_i) = gf' \zeta (3) T_{\rm hs,i}^3 / \pi^2$ and $\rho_Y(a_i) = gf \pi^2 T_{\rm hs,i}^4 / 30$, where $g$ is the degrees of freedom of the $Y$ particles, $f$ is 1 or 7/8 if the $Y$ particles are bosons or fermions respectively, and $f'$ is 1 if the $Y$ particles are bosons and $3/4$ if they are fermions. Equating the definitions of $\rho_Y(a_{\rm nr})$ from Eqs.~(\ref{rdev1}) and (\ref{rdev2}) and using the expressions for $n_Y(a_i)$ and $\rho_Y(a_i)$ from above with $a_p/a_i = bT_{\rm hs,i}/m$, we obtain \begin{equation}
    b = \frac{f}{f'} \frac{\pi^4}{30} \frac{1}{\zeta (3)}.
\end{equation} Substituting the values of $f$ and $f'$ yields $b = 2.70$ if the $Y$ particles are bosons and $b=3.15$ if they are fermions.

\begin{figure}[h!]
    \centering
    \includegraphics[width=\textwidth]{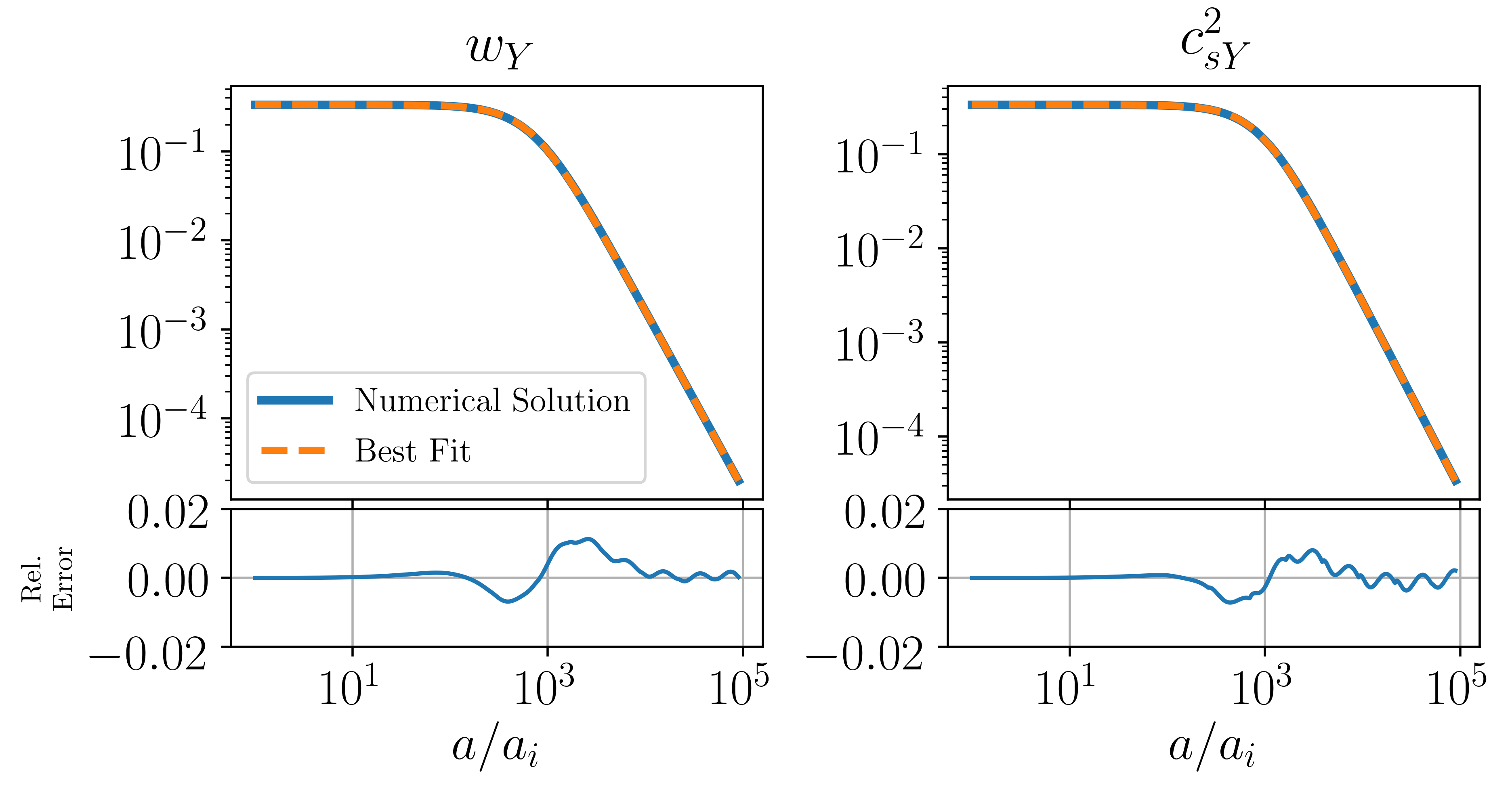} 
    \caption{\textbf{Top}: Numerical solutions and fitting functions for $w_Y$ (left) and $c_{sY}^2$ (right) for a fermion $Y$ particle of $m = 1$ TeV and $T_{\rm hs}(a_i) = 200m$. The fitting function curves use the best fit parameters given in Table \ref{wcst}. \textbf{Bottom}: The relative error between the numerical solution and the fit function, given by $1 - \rm Numerical / Fit$.} \label{fig-wcsfits}   
\end{figure}

The evolution of $w_Y$ and $c_{sY}^2$ can also be described by broken power laws. Both these quantities are equal to $1/3$ when the $Y$ particles are relativistic and are proportional to $a^{-2}$ when the $Y$ particles become nonrelativistic. This behavior is illustrated for a case with $m = 1$ TeV and $T_{\rm hs,i}= 200m$ by the blue solid curves in Fig.~\ref{fig-wcsfits}. We find that $w_Y$ and $c_{sY}^2$ are well-described by the functional form \begin{equation} \label{fitfunc}
    f(a,a_b,D) = \frac{1}{3} \left[ 1 + \left( \frac{a}{a_b} \right)^{\frac{1}{D}} \right] ^{-2 D},
\end{equation} where $a_b$ is the bending scale factor where the function transitions from the early-time power law to the late-time power law and $D$ models the width of the transition. 

Treating $a_b$ and $D$ as fit parameters, the numerical solutions for $w_Y$ and $c_{sY}^2$ were fit to the above function for a range of masses and values of $T_{\rm hs,i}/m$ for both boson and fermion $Y$ particles. The best fit values are presented in Table \ref{wcst}. The orange dashed lines in Figure~\ref{fig-wcsfits} show the functions of the form given by Eq.~(\ref{fitfunc}) with the best fit values of $a_b$ and $D$ given in Table \ref{wcst}. The bottom panels show the relative error between the numerical solution and the best fit functions. The error stays within 1.5\% for $w_Y$ and 1.2\% for $c_{sY}^2$ and stays within 0.2\% at late times for both quantities. 

We also used the fitting function for $w_Y$ with the best fit parameters and integrated Eq.~(\ref{rhoy-eq}) to obtain $\rho_Y$ to compare it with the numerical solution of $\rho_Y$. The relative error between the numerical and integrated $\rho_Y$ peaks at 0.5\% and stays constant at 0.2\% at late times. Our fitting forms for $w_Y$ can be used to obtain the time evolution of $\rho_Y$ to within this error.

\begin{table}[!h]
\centering  
\begin{tabular}{|c|cc|cc|}
\hline
Quantity    & \multicolumn{2}{c|}{$w_Y$}                               & \multicolumn{2}{c|}{$c_{sY}^2$}                          \\ \hline
            & \multicolumn{1}{c|}{$(a_b / a_i) / (T_{\rm hs,i}/m)$} & $D$ & \multicolumn{1}{c|}{$(a_b / a_i) / (T_{\rm hs,i}/m)$} & $D$ \\ \hline
Boson $Y$   & \multicolumn{1}{c|}{3.05}                    & 0.57     & \multicolumn{1}{c|}{3.91}                    & 0.55     \\ \hline
Fermion $Y$ & \multicolumn{1}{c|}{3.49}                    & 0.56     & \multicolumn{1}{c|}{4.48}                    & 0.54     \\ \hline
\end{tabular}
\caption{Best fit parameters for equation of state $w_Y$ and sound speed $c_{sY}^2$ for the functional form given by Eq.~(\ref{fitfunc}).}
\label{wcst}
\end{table}

In the derivation of the peak wavenumber in Sec.~\ref{peakscale}, we also use a piecewise model for $c_{sY}^2$, in which $c_{sY}^2$ is approximated as a sharply broken power law with a pivot point, so that \begin{equation}
    c_{sY}^2 = \begin{cases}
    \frac{1}{3},& a < a_{pc} , \\
    \frac{1}{3} \frac{a_{pc}^2}{a^2},& a>a_{pc},
    \end{cases}
\end{equation} with $a_{pc} = 1.43a_p$ for bosonic $Y$ particles and $a_{pc} = 1.41a_p$ for fermionic $Y$ particles, where $a_p$ is the pivot scale factor for the evolution of $\rho_Y(a)$.

\section{Relating The Start Of The EMDE To Model Parameters}
\label{impres}

In a universe that is initially dominated by relativistic SM particles, the EMDE starts when the energy density of the $Y$ particles exceeds the energy density of the SM particles at a scale factor $a_{\rm dom}$, when the SM temperature is $T_{\rm dom}$. Here, we derive a few important expressions for quantities related to the start of the EMDE in terms of the parameters of our model: $m$ (the $Y$ particle mass), $\eta$ (the ratio of the initial energy densities of SM radiation and the $Y$ particles), the reheat temperature $\rh{T}$ defined in Eq.~(\ref{gamma}), and $b$, which is 2.70 and 3.15 if the $Y$ particles are bosons and fermions, respectively. In the following, $g$ denotes the degrees of freedom of the $Y$ particles, and $f$ is 1 or 7/8 for boson or fermion $Y$ particles, respectively.

We first evaluate $a_{\rm dom}/a_p$, where $a_p$ is the pivot scale factor for the broken power law followed by $\rho_Y(a)$. Since entropy is conserved for the SM radiation before the $Y$ particle decays become significant, we have $\gs{T}a^3 T^3 = \rm constant$, where we assume $\gs{T} = g_{*S}(T)$. As a result, $\rho_R  \propto \gs{T(a)}^{-1/3} a^{-4}$. Then, \begin{equation} \label{rrb1} \frac{\rho_R(a_p)}{\rho_{\rm dom}} = \frac{g_{\rm dom}^{1/3} a_{\rm dom}^4}{g_p^{1/3} a_p^4}, \end{equation} where $g_{\rm dom} = \gs{T_{\rm dom}}$, $g_p = \gs{T(a_p)}$ and $\rho_{\rm dom} = \rho_R(a_{\rm dom}) = \rho_Y(a_{\rm dom}) = (\pi^2 /30) \gs{T_{\rm dom}} T_{\rm dom}^4$. Furthermore, since $\rho_Y \propto a^{-3}$ from $a_p$ to $a_{\rm dom}$, we can write \begin{equation} \label{rrb2}
    \frac{\rho_Y(a_p)}{\rho_{\rm dom}} = \frac{a_{\rm dom}^3}{a_p^3}.
\end{equation} Using Eqs.~(\ref{rrb1}) and (\ref{rrb2}), we have \begin{equation} \label{rrb3}
    \frac{\rho_R(a_p)}{\rho_Y(a_p)} = \left( \frac{g_{\rm dom}}{g_p} \right)^{\frac{1}{3}} \frac{a_{\rm dom}}{a_p}.
\end{equation} Similarly, we can express $\rho_R(a_p) = \rho_R(a_i) [g_i/g_p]^{1/3} [a_i/a_p]^4$, where $g_i = \gs{T_i}$. Since $\rho_Y \propto a^{-4}$ from $a_i$ to $a_p$, $\rho_Y(a_p) = \rho_Y(a_i) [a_i/a_p]^4 = \eta^{-1} \rho_R(a_i) [a_i/a_p]^4$. Combining the previous two expressions yields \begin{equation} \label{rrb4}
    \frac{\rho_R(a_p)}{\rho_Y(a_p)} = \eta \left( \frac{g_i}{g_p} \right)^{\frac{1}{3}}.
\end{equation} Equating Eqs.~(\ref{rrb3}) and (\ref{rrb4}) gives us \begin{equation} \label{adap}
    \frac{a_{\rm dom}}{a_p} = \left( \frac{g_i}{g_{\rm dom}} \right)^{\frac{1}{3}} \eta.
\end{equation} The above relation can be used to express $T_{\rm dom}$ in terms of our model parameters. We can use $\rho_Y \propto a^{-4}$ from $a_i$ to $a_p$ and $\rho_Y \propto a^{-3}$ from $a_p$ to $a_{\rm dom}$ to write \begin{equation} \label{ryadom}
    \rho_Y(a_{\rm dom}) = \rho_Y(a_i) \left( \frac{a_i}{a_p} \right)^4 \left( \frac{a_p}{a_{\rm dom}} \right)^3.
\end{equation} According to our model for the evolution of $\rho_Y(a)$, $a_p/a_i = bT_{\rm hs,i}/m$ and $\rho_Y(a_i) = (\pi^2 / 30) gf T_{\rm hs,i}^4$. In addition, we use the expression for $a_{\rm dom}/a_p$ from Eq.~(\ref{adap}) and equate $\rho_Y(a_{\rm dom})$ to $(\pi^2 /30) \gs{T_{\rm dom}} T_{\rm dom}^4$ to obtain \begin{equation}
    g_{\rm dom}^{\frac{1}{4}} T_{\rm dom} = (gf)^{\frac{1}{4}} \left( \frac{g_i}{g_{\rm dom}} \right)^{-\frac{1}{12}} \left( \frac{m}{b} \right) \eta^{- \frac{3}{4}},
\end{equation} which gives \begin{equation}\label{rhod}
    \rho_{\rm dom}  = gf \frac{\pi^2}{30} \left( \frac{g_i}{g_{\rm dom}} \right)^{-\frac{1}{3}} \left( \frac{m}{b} \right) ^4 \eta^{-3}.
\end{equation} 

Next, we derive an expression for $k_{\rm dom} / \rh{k} \equiv (aH)_{a_{\rm dom}} / (\rh{a} \Gamma)$. We divide Eq.~(\ref{adap}) by Eq.~(\ref{arhap}) to get $a_{\rm dom}/\rh{a}$, substitute $\Gamma$ from Eq.~(\ref{gamma}) and use $\rho_{\rm dom}$ from Eq.~(\ref{rhod}) in $H(a_{\rm dom}) = \sqrt{2 (8\pi G/3) \rho_{\rm dom}}$ to obtain \begin{equation}\label{kdom}
    \frac{k_{\mathrm{dom}}}{\rh{k}} = \sqrt{2} \left( \frac{gf}{\gs{\rh{T}}} \right)^{\frac{1}{6}} \left( \frac{\gs{T_i}}{\gs{T_{\rm dom}}} \right)^{\frac{1}{6}} \left( \frac{(m/b)}{\rh{T}} \right)^{\frac{2}{3}}  \eta^{-\frac{1}{2}}.
\end{equation} We also find it useful to derive an expression for $k/k_{\rm dom} = (aH)_{a_k} / (aH)_{a_{\rm dom}} $ for a mode $k$ that enters the horizon at $a_k$ during the period of radiation domination before the EMDE. Since the energy densities of the $Y$ particles and the radiation are equal at $a_{\rm dom}$, we have $H^2 (a_{\rm dom}) = 2 \times (8\pi G/3) \rho_R(a_{\rm dom})$. It follows that \begin{equation}
    \frac{H(a_k)} {H(a_{\rm dom})} = \frac{1}{\sqrt{2}} \left(\frac{g_{\rm dom}}{g_k} \right)^{\frac{1}{6}} \left(\frac{a_{\rm dom}}{a_k} \right)^2,
\end{equation} where $g_k = \gs{T(a_k)}$, and \begin{equation} \label{kkd}
\frac{k}{k_{\rm dom}} = \frac{1}{\sqrt{2}} \left(\frac{g_{\rm dom}}{g_k} \right)^{\frac{1}{6}} \frac{a_{\rm dom}}{a_k}.
\end{equation} Using Eq.~(\ref{kkd}) with Eq.~(\ref{adap}) yields \begin{equation} \label{kpkd}
    \frac{k_p}{k_{\rm dom}} =  \left(\frac{g_i^2}{g_p g_{\rm dom}}\right)^{\frac{1}{6}}\frac{\eta}{\sqrt{2}} 
\end{equation} for a universe with $\eta>1$. 

Finally, we obtain an expression for $k_y / k_{\rm dom} = a_y H(a_y) / (a_{\rm dom} H(a_{\rm dom}))$ where $a_y$ is the scale factor at which $T_{\rm hs} = m$. For this derivation, we relax the assumption of radiation domination before the EMDE because $\rho_Y$ contributes significantly to $H(a_y)$ for $\eta \lesssim 10$. Since the $Y$ particles are relativistic at $a_y$ and $T_{\rm hs} \propto a^{-1}$ for $a_i < a < a_y$, we can express $a_y = a_i T_{\rm hs,i}/m = a_p/b$. Using Eq.~(\ref{adap}), this yields
\begin{equation}\label{aday}
    \frac{a_{\rm dom}}{a_y} = \left( \frac{g_i}{g_{\rm dom}} \right)^{\frac{1}{3}} b\eta.
\end{equation} Next, we can express $H^2(a_y) = 8\pi G[\rho_Y(a_y) + \rho_R(a_y)]/3 = 8 \pi G \rho_R(a_y) [1 + \eta^{-1}] / 3 $. Using the scaling $\rho_R(a) \propto g(T(a))^{-1/3} a^{-4}$ with the definition of $\rho_{\rm dom}$ from Eq.~(\ref{rhod}) in $H^2(a_{\rm dom}) = 16 \pi G \rho_{\rm dom}/3$, and the expression for $a_{\rm dom}/a_y$ from Eq.~(\ref{aday}), we obtain \begin{equation} \label{kykd}
    \frac{k_y}{k_{\rm dom}} = \left(\frac{g_i^2}{g_y g_{\rm dom}}\right)^{\frac{1}{6}}\frac{b\eta \sqrt{1 + \eta^{-1}}}{\sqrt{2}}.
\end{equation}

\section{Perturbation Equations}
\label{pertsol}

We work in the Newtonian gauge:\begin{equation} ds^2 = -(1 + 2\psi)dt^2 + a^2(t)(1 + 2\phi)(dx^2 + dy^2 + dz^2). \end{equation} Ignoring anisotropic stress, we have $\psi = -\phi$. In the absence of decays, the general equations for the density contrast $\delta \equiv (\rho - \bar{\rho}) / \bar{\rho}$ and velocity dispersion $\theta \equiv a \partial_i dv^i / dt$ of a fluid for a Fourier mode $k$ are \cite{mabert95}: \begin{equation} \label{perteqs}
\begin{aligned}
     \delta ' + (1 + w) \frac{\theta}{aH}  + 3 (c_{s}^2 - w)\delta  + 3(1 + w) \phi '&= 0 \,, \\
\theta ' + (1 - 3w) \theta  + \frac{w'}{1+w} \theta -  k^2 \frac{c_{s}^2}{1 + w} \frac{\delta}{aH} + k^2 \frac{\phi}{aH} &= 0 \,,
\end{aligned}
\end{equation} where primes denote $d/d \ln a$, $w \equiv P/\rho$ is the ratio of the pressure and density of the fluid, and the sound speed is $c_{s}^2 \equiv \delta P / \delta \rho = P' / \rho' =  w - w'/3(1 + w)$.
    
The effects of $Y$ particles decaying into SM radiation are incorporated into the perturbation equations as in Ref. \cite{ae15}, which assumed a nonrelativistic $Y$ particle. Their treatment can be used because $w_Y$ is negligible in the epoch when the decay of the $Y$ particles is significant, i.e. $\Gamma w_Y / H \approx 0$ at all times. The $\mathcal{O} (\Gamma w_Y / H)$ corrections to these equations are given in Ref.~\cite{cannibal_big}. The full coupled system of equations for the three fluids and gravity is \begin{subequations} \label{perteq}
\begin{align}
\delta_X ' &= -\frac{\theta_X}{aH}  - 3 \phi ' \,; \\
\theta_X ' &= -  \theta_X  - k^2 \frac{\phi}{aH} \,; \label{thx} \\
\delta_Y ' &= -(1 + w_Y)\frac{\theta_Y}{aH}  - 3 (c_{sY}^2 - w_Y)\delta_Y - 3(1 + w_Y) \phi ' + \frac{\Gamma}{H}\phi \,; \label{dely}  \\
\theta_Y ' &= - (1 - 3w_Y) \theta_Y  - \frac{w'_Y}{1+w_Y} \theta_Y +  k^2 \frac{c_{sY}^2}{1 + w_Y} \frac{\delta_Y}{aH} - k^2 \frac{\phi}{aH} \,; \label{thy}\\
\delta_R ' &= -\frac{4}{3} \frac{\theta_R}{aH} - 4\phi ' +  \frac{\rho_Y}{\rho_R} \frac{\Gamma}{H} (\delta_Y - \delta_R -  \phi) \,; \\
\theta_R ' &= k^2 \frac{\delta_R}{4aH}  - k^2 \frac{\phi}{aH}  + \frac{\rho_Y}{\rho_R} \frac{\Gamma}{H} \left(\frac{3\theta_Y}{4} - \theta_R\right)\,; \\
\phi ' &= - \left( 1 + \frac{k^2}{3H^2 a^2} \right) \phi + \frac{4 \pi G}{3 H^2} \left(\sum \delta_i \rho_i\right) \label{poisson}.
\end{align}
\end{subequations} 

To determine the initial conditions of the system given by Eqs.~(\ref{perteq}), we first set $\phi(a=a_i) = \Phi_0$. We assume adiabatic perturbations and equate the primordial curvature perturbation for all three species: \begin{equation} \label{zetai}
    \zeta_j = \Phi - \frac{\delta_j}{[\ln \rho_j]'},
\end{equation} where $j$ indicates each fluid. For superhorizon modes $\zeta = 3\Phi_0/2$ in a universe dominated by radiation or relativistic $Y$ particles. Setting $\zeta_j = \zeta$ for each species, we have the initial conditions, \begin{subequations}
\begin{align}
    \frac{\delta_R}{\Phi_0} &= 2 \,; \\
    \frac{\delta_X}{\Phi_0} &= \frac{3}{2} \,; \\ 
    \frac{\delta_Y}{\Phi_0} &= -\frac{1}{2} [\ln \rho_Y ] ' |_{a = a_i},
\end{align} where $a_i$ is the scale factor at which our calculations begin, chosen such that $T_{\rm hs}(a_i) = 300m$.
\end{subequations} The initial conditions for the velocity dispersions are \cite{mabert95} \begin{equation} \label{thetai}
   \theta_R =  \theta_X =  \theta_Y = -\frac{k^2 \Phi_0}{2H(a_i)}.
\end{equation}

For a universe that is initially dominated by nonrelativistic $Y$ particles, the initial conditions are similar to those in matter domination. For superhorizon modes, the primordial curvature perturbation is related to the metric perturbation as $\zeta_0 = 5 \Phi_{0m}/3$. The primordial curvature perturbations for all species, given by Eq.~(\ref{zetai}), are set equal to each other, yielding \begin{subequations}
\begin{align}
    \frac{\delta_R}{\Phi_{0m}} &= 8/3 \,; \\
    \frac{\delta_X}{\Phi_{0m}} &= 2 \,; \\ 
    \frac{\delta_Y}{\Phi_{0m}} &= 2.
\end{align} The initial conditions for $\theta_i$ are given by Eq.~(\ref{thetai}) with $\Phi_{0m}$ replacing $\Phi_0$.
\end{subequations}

\label{ydom-pert}

\section{EMDE Power Spectrum Application}
\label{app}

The EMDE modifies the matter power spectrum for modes that enter the horizon during or before the EMDE ($k > \rh{k}$). For an EMDE that results from cold $Y$ particles dominating the universe after inflation, this modification to the power spectrum was described by Ref. \cite{emde2}. For $k < 0.05 \rh{k}$, the power spectrum remains the same. For $k > 0.05 \rh{k}$, $\delta(k) \rightarrow R(k) \delta(k)$, where \begin{equation}
    R(k) = \frac{A\left(\frac{k}{0.86\rh{k}}\right) \ln \left[\left(\frac{4}{e^3}\right)^\frac{f_2}{f_1}\frac{B\left(\frac{k}{0.86\rh{k}}\right) a_\mathrm{eq}}{a_k}\right]}{9.11 \ln \left[\left(\frac{4}{e^3}\right)^\frac{f_2}{f_1} 0.594  \frac{\sqrt{2} k}{k_\mathrm{eq}}\right]}.
\end{equation} In this equation, $a_k$ is the scale factor of horizon entry for mode $k$ and $a_{\rm eq}$ and $k_{\rm eq}$ are the scale factor and horizon wavenumber at matter-radiation equality, respectively. The values of $f_1$ and $f_2$ are determined by the baryon fraction $f_\mathrm{b} \equiv \rho_\mathrm{bar} /(\rho_\mathrm{b}+\rho_{\rm matter})$:
\beqa
f_1 &=& 1 - 0.568 f_\mathrm{b} + 0.094 f_\mathrm{b}^2 \nonumber;\\
f_2 &=& 1 - 1.156 f_\mathrm{b} + 0.149 f_\mathrm{b}^2 - 0.074 f_\mathrm{b}^3 \nonumber.
\eeqa Furthermore, \begin{equation}
    \frac{a_\mathrm{eq}}{a_k} = \frac{\sqrt{2} k}{k_\mathrm{eq}}\left[1+\left(\frac{k}{k_\mathrm{RH}}\right)^{4.235}\right]^{1/4.235},
\end{equation} and the fitting functions for $A$ and $B$ are: \beqa
A(x) &=& \exp\left[\frac{0.609}{\{1+2.15(\ln x -1.52)^2\}^{1.38}}\right] \\
&&\times \left[9.11 \,{\cal S}(5.02-x) +\frac{3}{5}x^2 \, \,{\cal S}(x-5.02)\right];\nonumber\\
\ln B(x) &=& \ln(0.594)  \,{\cal S}(5.02-x) + \ln\left(\frac{e}{x^2}\right)  \,{\cal S}(x-5.02),\nonumber
\eeqa
where
\begin{equation}
    {\cal S}(y) = \frac{1}{2}\left[ \tanh \left(\frac{y}{2}\right) +1\right]
\end{equation} models a step function.

If an epoch of SM radiation domination precedes the EMDE, modes with $k > k_{\rm dom}$ grow logarithmically with scale factor after entering the horizon and before the EMDE. This modifies $\delta(k > k_{\rm dom})$. From our fitting function for $\delta_{Y,c}(k > 10 \rh{k})$ given by Eq.~(\ref{dyc-yd}), we find that this modification is modeled by the scale-dependent factor \begin{equation}
    R_{\rm RD} (q) = \frac{\ln (1 + 0.22q)}{0.22q}[1 + 1.11q  + (0.94q)^2 +  (0.63q)^3 + (0.45q)^4]^{-1/4},
\end{equation} where $q = k/k_{\rm dom}$ and $R_{\rm RD} = 1$ for $\eta < 1$. Finally, the small-scale cut-off can be imposed on $\delta(k)$ using our transfer functions $T(k) = \exp [-(k/k_{\rm cut})^n]$ from Section \ref{tf}. In summary, the combined effect of the EMDE, an epoch of radiation domination before the EMDE, and the small-scale cut-off due to the relativistic pressure of the $Y$ particles modifies $\delta(k)$ by a factor \begin{equation}
    R_{\rm EMD} (k) = R(k) R_{\rm RD} (k) T(k).
\end{equation}

The above expression for $R_{\rm EMD}(k)$ is valid at all times after the EMDE ends. We provide an online application for the easy computation and visualization of $R_{\rm EMD}$\footnote{\href{https://hganjoo-emde-emde-rk-s7ww2v.streamlitapp.com/}{{https://hganjoo-emde-emde-rk-s7ww2v.streamlitapp.com/}}}. The calculations of the peak and cut-off scales in the application neglect the variation of $g_{*}$, the number of relativistic degrees of freedom in the SM radiation, before the EMDE. The parameters $\rh{T}$, $\eta$ and $k_{\rm dom} / \rh{k}$ or $k_y / \rh{k}$ can be varied by the user, and the output is downloadable as a table.

\bibliography{refs.bib}

\end{document}